\documentclass[12pt]{article}

\usepackage{amsmath,amsfonts,amssymb,color,graphicx,overpic,} 
\usepackage{slashed}
\usepackage{epsfig}
\usepackage{amsmath}
\usepackage{ wasysym }
\usepackage{graphicx}
\usepackage[english]{babel}
\usepackage{graphic x}
\usepackage{relsize}

\begin{document}


\renewcommand{\arraystretch}{2}

\begin{titlepage}
\rightline{\large September 2014}
\vskip 2cm
\centerline{\Large \bf
Dissipative hidden sector dark matter}

\vskip 2.2cm
\centerline{\large R.Foot\footnote{E-mail address:
rfoot@unimelb.edu.au}, S.Vagnozzi\footnote{
E-mail address: svagnozzi@physics.unimelb.edu.au, vagnozzi@nbi.dk}}

\vskip 0.7cm
\centerline{\it ARC Centre of Excellence for Particle Physics at the
Terascale,}
\centerline{\it School of Physics, University of Melbourne,}
\centerline{\it Victoria 3010 Australia}
\vskip 2cm
\noindent

A simple way of explaining dark matter without modifying known Standard
Model physics is to require the existence of a hidden (dark) sector, 
which interacts with the visible one predominantly via gravity. We
consider a hidden sector containing two stable particles charged 
under an unbroken $U(1)^{'}$ gauge symmetry, hence featuring dissipative
interactions. The massless gauge field associated with 
this symmetry, the dark photon, can interact via kinetic mixing with the
ordinary photon. In fact, such an interaction of strength $\epsilon \sim
10 ^{-9}$ appears to 
be necessary in order to explain galactic structure. We calculate the
effect of this new physics on Big Bang Nucleosynthesis and its 
contribution to the relativistic energy density at Hydrogen
recombination. 
We then examine the process of dark recombination, during which neutral
dark states are formed, which is important for large-scale structure
formation. 
Galactic structure is considered next, focussing on spiral and irregular
galaxies. For these galaxies we modelled the dark matter halo (at the
current epoch) as
a dissipative plasma of dark matter particles, where the energy lost due
to dissipation is compensated by the energy
produced from ordinary supernovae (the core-collapse energy is
transferred to the hidden sector via kinetic mixing induced
processes in the supernova core).  
We find that such a dynamical halo model can reproduce several observed 
features of disk galaxies, including the cored density profile and the
Tully-Fisher relation. 
We also discuss how elliptical and dwarf spheroidal galaxies could fit
into this picture.
Finally, these analyses are combined to set bounds on the parameter
space of our model, which can serve as a guideline for future
experimental searches.

 \end{titlepage}
 
 \newpage

\section{Introduction}

A variety of observations suggest the existence of non-baryonic dark
matter in the Universe. Among these are measurements of the rotation
curves of spiral galaxies, which are asymptotically flat \cite{rubin}.
Dark matter is also required to explain the Cosmic Microwave Background
(CMB) anisotropy spectrum (particularly the structure of the acoustic
peaks), the matter power spectrum and large-scale structure (LSS)
formation (see e.g. \cite{refregier}). Cosmological observations can be
explained within the framework of the Friedmann-Robertson-Walker (FRW)
model (see e.g. \cite{dodelson}), which assumes isotropy and homogeneity
of the Universe on large scales. Comparison with observations require
the total dark matter mass to be approximately five times that of
baryonic matter.

The particle physics underlying dark matter is unknown but a promising
possibility, widely discussed in recent literature (see e.g.
\cite{feng,cline6,ringwald,broken}) is that dark matter resides in a
hidden sector. That is, an additional sector containing particles and
forces which interact with the known Standard Model particle content
predominantly via gravity. A special case is mirror dark matter (MDM),
where the hidden sector is exactly isomorphic to the Standard Model
\cite{fundamentalimproper}. It has been shown that MDM can, under
suitable assumptions and initial conditions, reproduce the successes of
collisionless cold dark matter (CDM) on large scales, while deviating on
small scales. This is important because such a model has the potential
to address apparent shortcomings of collisionless CDM such as inferred
cores in dark matter halos and the missing satellites problem
\cite{berezhiani,mirrormattertype}.

Mirror dark matter is self-interacting due to an unbroken $U(1) ^{'}$
interaction (mirror electromagnetism). The associated gauge boson, the
mirror photon, is massless, which implies that MDM is dissipative.
Dissipative dark matter is a possible scenario, provided that there
exists a substantial heat source that can replace the energy lost due to
dissipative interactions. It has been argued \cite{spheroidal} that
ordinary supernovae can provide such a heat source provided
photon-mirror photon kinetic mixing exists. More in-depth studies of
this possibility \cite{depth4} have shown that the model can reproduce
several observational properties of disk galaxies. MDM also seems to be
capable of explaining the positive results from the direct detection
experiments, especially the annual modulation signals observed by DAMA
\cite{dama1} and CoGeNT \cite{cogent}, consistently with results from
the other experiments \cite{dama,electronscattering}. For an up-to-date
review and more detailed bibliography see \cite{review}.

It is possible that dark matter might arise from a more generic hidden
sector with qualitatively similar features. So long as the hidden sector
contains an
unbroken $U(1) ^{'}$ gauge interaction, dissipative dark matter can
arise. The simplest such generic hidden sector model contains two
massive states,
interacting with the $U(1) ^{'}$ gauge field (the \textit{dark photon}),
with a priori unknown $U(1) ^{'}$ charges and masses. Such a model can
then closely
resemble MDM, with the lighter state corresponding to the mirror
electron and the heavier state corresponding to mirror nuclei. Kinetic
mixing can couple the
massless $U(1) ^{'}$ gauge field with the ordinary photon. The
fundamental physics is 
described by five free parameters. Our aim is to constrain this
5-dimensional parameter space using early Universe cosmology and
galactic structure considerations.

The outline of this article, then, will be as follows. In Section 2 we
define the model and examine some of its properties. Sections 3 and 4
will be devoted
to studying its early Universe phenomenology, focussing in particular on
how Big Bang Nucleosynthesis (BBN) and the onset of structure formation
are affected. Section 5 is dedicated to analyzing the model in the
context of galactic structure. Finally, in Section 6 we draw on the
analyses of the previous sections to summarize the constraints on the
model and in Section 7 we give some concluding remarks.

\section{Two-component hidden sector dark matter}

The model considered incorporates a hidden sector featuring an unbroken
$U(1) ^{'}$ gauge interaction. 
This means there is a massless gauge boson, called the
\textit{dark photon} ($\gamma _{_D}$). The hidden sector will also
contain two stable dark matter particles, $F_1$ and $F_2$, taken to be
Dirac fermions, with
masses $m _{F_1}$ and $m _{F_2}$. These two particles are assumed to be
charged under the $U(1) ^{'}$ gauge group, with charges $Q _{F_1} ^{'}$
and $Q _{F_2}
^{'}$, opposite in sign but not necessarily equal in magnitude.

In the early Universe, the $U(1)'$ interactions would be expected to
efficiently annihilate the symmetric component, meaning that the
abundance of $F_1$ and $F_2$ dark matter is set by its
particle-antiparticle asymmetry. This is an example of
\textit{asymmetric} dark matter, which has been extensively discussed in
recent literature \cite{reviewadm}. Dark matter asymmetry and local
neutrality of the Universe then imply:
\begin{eqnarray}
n _{F_1}Q _{F_1} ^{'} + n _{F_2}Q _{F_2} ^{'} = 0 \ ,
\end{eqnarray}
where $n _{F_1}$ and $n _{F_2}$ are the number densities of $F _1$ and
$F _2$ respectively. This is, of course, quite analogous to the
situation with ordinary matter ($F_1 \sim$ electron, $F_2 \sim$ proton).

The only possible renormalizable and gauge-invariant interaction 
coupling the ordinary particles with the dark sector is the $U(1)' -
U(1)_Y$ kinetic mixing
term \cite{foothe}. Including this term, the 
the full Lagrangian of our model is:\footnote{Here and throughout the
article, natural units with $\hbar = c = k _B = 1$ will be used.}
\begin{eqnarray}
{\cal L} = {\cal L} _{\text{SM}} -\frac{1}{4}F ^{'\mu \nu} F _{\mu \nu}
^{'} + \overline{F} _1(iD _{\mu}\gamma ^{\mu} - m _{F_1})F _1 +
\overline{F} _2(iD _{\mu}\gamma ^{\mu} - m _{F_2})F _2 - \frac{\epsilon
^{'}}{2}F ^{\mu \nu} F _{\mu \nu} ^{'} \ ,
\end{eqnarray}
where ${\cal L}_{SM}$ denotes the $SU(3)_c \otimes SU(2)_L \otimes
U(1)_Y$ gauge invariant Standard Model Lagrangian which describes
the interactions of the ordinary particles.
Also, $F _{\mu \nu} ^{'} = \partial _{\mu} A _{\nu} ^{'} - \partial
_{\nu} A _{\mu} ^{'}$ 
[$F _{\mu \nu}  = \partial _{\mu} B _{\nu}  - \partial _{\nu} B _{\mu}
$] 
is the field-strength tensor 
associated with the $U(1) ^{'}$ [$U(1)_Y$] gauge interaction, $A _{\mu}
^{'}$ [$B_{\mu} = \cos\theta_w A_\mu + \sin\theta_w Z_\mu$]
being the relevant gauge field. 
The two dark fermions are described by the quantum fields $F_j$ and $D
_{\mu}F _j = \partial _{\mu}F _j + ig ^{'}Q _{j} ^{'} A _{\mu} ^{'}F
_j$, where $g ^{'}$ is the coupling constant relevant to this gauge
interaction ($j=1,2$). The dark fermions are stable which is a
consequence of the $U(1)'$
gauge symmetry and an accidental $U(1)$ global symmetry  (implying
conservation of $F_1$ and $F_2$ number). This is reminiscent of how
$U(1)_Q$ and the accidental baryon number  symmetries arise in the
Standard Model and how they stabilize the electron and proton. This is
quite a general feature of hidden sector dark matter models and
illustrates why they are so appealing theoretically: they typically
predict a spectrum of massive, dark and stable particles.

The interactions of $F _1$ with the dark photon are characterized by the
dark fine structure constant: $\alpha ^{'} \equiv (g'Q _{F_1}) ^2/4\pi$.
The coupling of $F _2$ with the dark photon will be modified by the
charge ratio: $Z' \equiv Q _{F_2} ^{'}/Q _{F_1} ^{'}$. By means of a
non-orthogonal transformation, one can remove the kinetic mixing and
show that the net effect of the relevant term is to provide the dark
fermions with a tiny ordinary electric charge \cite{holdom}. The
physical photon now couples to dark fermions with charge:
\begin{eqnarray}
\cos\theta_w g'Q _{F_j} ^{'} \epsilon ^{'} \equiv \epsilon _{F_j}e \ .
\end{eqnarray}
Thus the fundamental physics of the model is described by 5 independent
parameters: $m _{F_1},m _{F_2},\alpha ^{'},Z ^{'}$ and $\epsilon \equiv
\epsilon _{F_1}$ (note that $\epsilon _{F_2} = Z ^{'}\epsilon _{F_1}$,
and is therefore not an independent parameter). For definiteness we will
focus on the case $m _{F_1} \ll m _{F_2}$, with $Z ^{'}$ being an
integer.

Clearly it is entirely possible for our model to be the low energy
effective field theory limit of a more complex theory. In this context,
$F _1$ and/or $F _2$ might represent bound states (dark nuclei), which
could be bound together by some interaction which resembles the strong
one. In this case, the $F_1$/$F_2$ masses arise from a dark confinement
scale (analogous to $\Lambda _{\text{QCD}}$ in the Standard Model)
rather than being a bare mass term. Alternatively, the mass terms for $F
_j$ might originate from a hidden sector scalar, $S$, by means of a
Lagrangian term of the form $\lambda _jS\bar{F} _jF _j$, with $\langle S
\rangle \neq 0$.

Some possible implications of this model for dark matter direct
detection experiments have been discussed previously in
\cite{hiddensector}. Furthermore, as explained in the introduction, the
dark matter phenomenology is similar but generalizes the MDM case.
Related hidden sector models, featuring an unbroken $U(1) ^{'}$
interaction, have also been discussed in recent literature, e.g.
\cite{feng,cline6,petrakiatomic} and much earlier in
\cite{goldberghall}. However, these models assume parameter space where
the dark matter galactic halo is in the form of atoms (or a non
dissipative plasma), and thus can be collisional but generally not
dissipative.\footnote{An alternative possibility examined in recent
literature, known as Double-Disk Dark Matter (DDDM), explores the
scenario where only a subdominant component of the dark matter exhibits
dissipative interactions \cite{dddm}. These dissipative dynamics allow
for DDDM to cool efficiently and form a thin dark matter disk, similar
to the baryonic disk.} We consider the case where the galactic halo is
in the form of a roughly spherical dissipative plasma. Such a spherical
plasma would cool via dissipative processes, for instance dark
bremsstrahlung, unless a substantial heat source exists. Here a pivotal
role is played by the 
kinetic mixing interaction: kinetic mixing induced processes 
(such as plasmon decay \cite{raf,updated}) 
within the core of ordinary core-collapse supernovae 
are presumed to provide the heat source that replaces the energy lost to
dissipative interactions. This is possible provided $\epsilon \sim 10
^{-9}$ and $m _{F_1} \lesssim {\rm few} \times T _{\text{SN}} \simeq 100
\ {\rm MeV}$, where $T _{\text{SN}}$ is the temperature reached in the
core of ordinary supernovae.\footnote{Although the kinetic mixing
parameter is very small, $\epsilon \sim 10 ^{-9}$, this does not
represent a theoretical problem, such as radiative instability. Indeed,
as discussed in \cite{poincare}, small values for the coupling are
technically natural (in the sense of 't Hooft \cite{hooft}) since, in the limit
$\epsilon \rightarrow 0$, an enhanced Poincar\'e symmetry arises: ${\cal
G} _P ^{\text{SM}} \otimes {\cal G} _P ^{\text{HS}}$, where ${\cal G}
_P$ denotes the Poincar\'e group and SM and HS stand for Standard Model
and Hidden Sector respectively.} A lower limit $m _{F_1} \gtrsim 0.01 \
{\rm MeV}$ arises from studies of Red Giants \cite{redgiants} and White
Dwarfs \cite{whitedwarfs,updated} (see \cite{bailey} for a summary of
relevant bounds).

Finally, one can also consider a two-component hidden sector model where
the two dark matter particles are bosons rather than fermions, 
charged under a $U(1)'$ gauge interaction. In the case of two scalar
particles, $B _j$, the Lagrangian is:
\begin{eqnarray}
{\cal L} = {\cal L} _{\text{SM}} - \frac{1}{4}F ^{'\mu \nu}F ^{'} _{\mu
\nu} + \left ( D _{\mu}B _j \right ) ^{\dagger}\left ( D ^{\mu}B _j
\right ) - m _{B _j} ^2B _j ^{\dagger}B _j - \frac{\epsilon ^{'} _B}{2}F
^{\mu \nu} F _{\mu \nu} ^{'} \ ,
\label{bosonic}
\end{eqnarray}
where $j=1,2$ (with summation over $j$ implied). As in the two-component
fermion case, one can consider the general case where the $U(1)'$
charges of $B _1$
and $B _2$ are different, with $Z' _B$ being the charge ratio. As
before, the Lagrangian in Eq.(\ref{bosonic}) possesses an accidental
global $U(1)$ 
symmetry which together with the $U(1)'$ gauge symmetry 
implies conservation of $B _1$ and $B _2$ number, and hence stability of
the dark matter particles. Again, the kinetic mixing term will play a 
dominant role in the cosmological and galactical dynamics of such a
model. Barring factors of order unity to account for spin statistics,
the analysis we will perform in the following sections will hold for
this bosonic model as well as the fermionic one. In particular, the
bounds that are summarized in Section 6 hold for the bosonic case, up to
factors of order unity. For definiteness, though, we will focus on the
fermionic model.

\section{Cosmology of the early Universe}
\vskip 0.5cm

In Sections 3 and 4 we derive constraints on the parameter space of the
model from early Universe cosmology considerations. 
We assume at the outset that the light $F_1$ particle has mass in the
range:
$0.01 \ {\rm MeV} \lesssim m _{F_1} \lesssim 100 \ {\rm MeV}$.
As mentioned in the previous section, and discussed in more detail in
Section 5, this mass range for the $F_1$ particle is 
motivated by the adopted dissipative dynamics governing galactic halos,
which sees 
substantial halo heating from ordinary core-collapse supernovae  
compensating for the energy lost due to dissipative interactions.
This mechanism also requires kinetic mixing of magnitude $\epsilon \sim
10^{-9}$ which, it turns out, is in the interesting
range where it can be probed by early Universe cosmology.

\vskip 0.3cm

\subsection{Evolution of $\frac{T _{\gamma _{_D}}}{T _{\gamma}}$}
\vskip 0.1cm

Successful cosmology, BBN and LSS strongly constrain exotic
contributions to the energy density during the radiation dominated era.
If we define $T _{\gamma}$[$T _{\gamma _{_D}}$] and $T _{\nu}$ to be the
photon [dark photon] and neutrino temperatures, then we require $T
_{\gamma _{_D}} \ll T _{\gamma}$ (the exact mechanism that provides such
an initial condition will not be of our concern, although asymmetric
reheating is possible within inflationary models
\cite{hodges}).\footnote{We only require $T _{\gamma _{_D}} \ll T
_{\gamma}$ at, say, the QCD phase transition, $T _{\text{QCD}} \sim 100$
MeV. Thus, even if the Universe started with $T _{\gamma _{_D}} = T
_{\gamma}$ at $T > T _{\text{QCD}}$, the heating of the ordinary sector
at the QCD phase transition would be sufficient to establish the
necessary initial condition, $T _{\gamma _{_D}} \ll T _{\gamma}$, at $T
_{\text{QCD}}$.} As discussed previously, kinetic mixing confers a tiny
ordinary electric charge to dark fermions. It follows that in the early
Universe energy and entropy can be transferred between the sectors. Thus
even if the Universe starts with $T _{\gamma _{_D}}/T _{\gamma}=0$, $T
_{\gamma _{_D}}$ will be generated as entropy is transferred from the
visible to the hidden sector. In the following work we first study the
evolution of $T _{\gamma _{_D}}/T _{\gamma}$ (with initial condition $T
_{\gamma _{_D}}/T _{\gamma} = 0$), and then consider the relevant
cosmological constraints.

In the early Universe energy is transferred between the sectors via
various processes, including (to order $\epsilon ^2$) $e \overline{e}
\rightarrow F _1 \overline{F} _1$, $eF_1 \rightarrow eF_1$, $e
\overline{e} \rightarrow F _2 \overline{F} _2$, $\gamma F_1 \rightarrow
\gamma _{_D}F _1$, and so on. Given the assumed initial condition, $T
_{\gamma _{_D}}/T _{\gamma} = 0$, we can to a reasonable approximation
neglect inverse processes, such as $F_1\overline{F}_1 \rightarrow
e\overline{e}$. 
Also, processes involving $F_2$ can be approximately neglected if $F_2$
is much heavier than $F_1$ [simple analytic calculations indicate that
for $m _{F_2} \gg {Z'} ^2\max (m _e , m _{F_1})$ the energy transfer
between the sectors is dominated by $F_1$ production]. Of the remaining
processes, $e \overline{e} \rightarrow F _1 \overline{F} _1$ is expected
to dominate (for $T _{\gamma} \gtrsim m _e$), given that the rates of
all other two-body processes are smaller by a factor of $\lesssim n
_{F_1}/n _e \sim (T _{\gamma _{_D}}/T _{\gamma}) ^3$, and typically we
are constrained to reside in the region of parameter space where $(T
_{\gamma _{_D}}/T _{\gamma}) ^3 \ll 1$. Hence for $m _{F_1} \gtrsim 0.1
\ {\rm MeV}$, we consider just one production process, $e\overline{e}
\rightarrow F_1\overline{F}_1$.\footnote{While this work was in
progress, the paper \cite{redondo} appeared which considered
$N_{\text{eff}}$ constraints on a related model. There they considered
additional production channels, such as $\gamma F_1 \rightarrow
\gamma_{_D} F_1$, for a wide range of parameter space. The effect of
these extra channels is to tighten constraints on $\epsilon$ by around a
factor of 2.}

For $m _{F_1} \lesssim 0.1$ MeV, one could consider  processes such as
$\gamma F _1 \rightarrow \gamma _{_D}F _1$ in addition to $e\overline{e}
\rightarrow F _1\overline{F} _1$. Although the rate for $\gamma F_1
\rightarrow \gamma _{_D}F _1$ is suppressed relative to $e\overline{e}
\rightarrow F_1 \overline{F} _1$ by $\sim \left ( T _{\gamma _{_D}}/T
_{\gamma} \right ) ^3$ for $T _{\gamma} \gtrsim m _e$, for $T _{\gamma}
\lesssim m _e$ the rate of $\gamma F_1 \rightarrow \gamma _{_D}F _1$ can
become important and eventually dominate.\footnote{Another $F_1$
production channel that could be relevant for very low $F_1$ mass is
plasmon decay ($\gamma \rightarrow F _1\overline{F} _1$). It can become
important when $m _{F_1} \lesssim \omega _P/2$, where $\omega _P =
\sqrt{4\pi\alpha T ^2/9}$ is the plasma frequency (see e.g.
\cite{updated}). This implies that during the period of interest (from
BBN to the formation of the CMB) plasmon decay is only important for $m
_{F_1} \lesssim 50 \ {\rm keV}$.} Here we shall focus on $m _{F_1}
\gtrsim 0.1$ MeV, where $e\overline{e} \rightarrow F _1\overline{F} _1$
is the dominant process affecting the evolution of the temperatures.
Thus, our analysis will only be strictly valid in the range $0.1 \ {\rm
MeV} \lesssim m _{F_1} \lesssim 100 \ {\rm MeV}$, while the study of the
region $0.01 \ {\rm MeV} \lesssim m _{F_1} \lesssim 0.1 \ {\rm MeV}$
will require further work. Restricting ourselves to the region of
parameter space $m _{F_1} \gtrsim 0.1 \ {\rm MeV}$ also bypasses several
other complications which arise in the context of galactic structure
(Section 5).

The cross-section for $e \overline{e} \rightarrow F_1 \overline{F}_1$ is
analogous to that of muon pair-production, with the essential difference
being that the coupling of $F_1$ to the ordinary photon is now given by
$\epsilon e$. The cross-section for this process is:
\begin{eqnarray}
\sigma = \frac{4\pi}{3s^3} \epsilon ^2 \alpha ^2 \sqrt{\frac{s - 4 m
_{F_1} ^2}{s - 4 m _{e} ^2}} \left[ 
s^2 + 2 \left (m _{e} ^2 + m _{F_1} ^2 \right )s + 4m _{e} ^2 m _{F_1}
^2 \right] \ ,
\end{eqnarray}
where $\sqrt{s}$ is the centre-of-momentum energy of the system, $\alpha
= e^2/4\pi$ is the fine-structure constant and $m _{e}$ is the electron
mass. The following treatment generalizes the MDM case analyzed in
\cite{predictions}, which itself followed earlier works
\cite{updated,carlson,ciarcellutiliege}. Energy is transferred between
the visible and dark sectors within a co-moving volume $R^3$ ($R$ being
the scale factor) at a rate given by:
\begin{eqnarray}
\frac{dQ}{dt} = R^3 n _{e} n _{\overline{e}} \langle \sigma v_{\text{M\o
l}} {\cal E} \rangle \ ,
\label{dqdt}
\end{eqnarray}
where $\langle \sigma v_{\text{M\o  l}} {\cal E}\rangle$ denotes the thermal
average of the cross-section ($\sigma$) the M\o ller velocity
($v_{\text{M\o l}}$) and the total energy of the process (${\cal E} = E_1 + E_2$).
Following \cite{ciarcellutiliege,gondologelmini}, we replace the exact
Fermi-Dirac distribution 
with the simpler Maxwellian one, so that the thermally averaged
cross-section is given by:
\begin{eqnarray}
\langle \sigma v_{\text{M\o l}} {\cal E} \rangle = \frac{\mathlarger
\int d^3p_1 d^3p_2 \ e^{-\frac{E_1}{T}} e^{-\frac{E_2}{T}} \ \sigma
v_{\text{M\o l}}{\cal E}}{\mathlarger \int d^3p_1 d^3p_2 \
e^{-\frac{E_1}{T}}e^{-\frac{E_2}{T}}} \ .
\end{eqnarray}
To evaluate the thermally averaged cross-section, similar steps as in
\cite{ciarcellutiliege,gondologelmini} can be followed, yielding:
\begin{eqnarray}
\langle \sigma v_{\text{M\o l}}{\cal E} \rangle = 
\frac{\omega}{8m_e^4 T_{\gamma}^2 {[K_2(\frac{m_e}{T_{\gamma}})]}^2}
\mathlarger \int _{4{\cal M}^2}^{\infty} ds \ \sigma (s - 4m_e^2)
\sqrt{s} \mathlarger \int _{\sqrt{s}}^{\infty} dE_+ \ e^{-\frac{E_+}{T
_{\gamma}}} E_+ \sqrt{\frac{E _+ ^2}{s} - 1} \ ,
\label{sig}
\end{eqnarray}
where $\omega \approx 0.8$ takes into account various approximations
such as the aforementioned use of a Maxwell-Boltzmann distribution in
lieu of the actual Fermi-Dirac one in evaluating the thermally averaged
cross-section \cite{ciarcellutiliege}. $K_2(z)$ is the modified Bessel
function of the second kind and argument $z$, and ${\cal M} \equiv
\max(m _e, m _{F_1})$. Finally, we can write (see for instance
\cite{earlyuniverse}):
\begin{eqnarray}
n _{e} \simeq n _{\overline{e}} = \frac{1}{\pi ^2} \int _{m_e} ^{\infty}
dE \ \frac{\sqrt{E^2 - m _e ^2}E}{1 + e^{\frac{E}{T _{\gamma}}}} \ .
\end{eqnarray}

Since self-interaction rates are bigger than the rates of kinetic mixing
induced processes by many orders of magnitude ($\sim 1/\epsilon ^2$),
the overall
system can be modelled as being composed of two subsystems, one at
temperature $T _{\gamma}$ and the other at temperature $T _{\gamma
_{_D}}$, exchanging
energy while remaining instantaneously in thermodynamical equilibrium.
This system is somewhat analogous to that of a block of ice melting in a
glass of water
(e.g. \cite{feynman}). The second law of thermodynamics can therefore be
applied to it. In principle, the neutrino subsystem should be taken into
account too.
In practice the net transfer of energy to the neutrino subsystem can be
approximately neglected, at least for $m _{F_1} \lesssim 10 \ {\rm
MeV}$, since energy
transfer to the dark sector then happens predominantly after neutrino
kinetic decoupling.\footnote{For $F_1$ masses in the range $10 \ {\rm
MeV} \lesssim m _{F_1} \lesssim 100 \ {\rm MeV}$, there can be
significant transfer of entropy out of 
the neutrino subsystem. For the largest $F_1$ masses, $m_{F_1} \sim 100$
MeV, the evolution can be separated into 
two distinct stages. The first is where $F_1, \bar F_1$ states are
produced via processes such as $\bar e e \to \bar F_1 F_1$.  
For these largest $F_1$ masses of interest, these production processes
will only be important for 
temperatures above the kinetic decoupling of the neutrinos so that
$T_\nu = T_\gamma$ results. The second stage is the annihilation
of electrons and positrons ($\bar e e \to \gamma \gamma$) which
continues to occur at temperatures where the neutrinos have kinetically
decoupled and leads
to the heating of photons relative to the neutrinos ($T_\gamma >
T_\nu$). 
We have checked that
the effect of neglecting the transfer of entropy to the neutrino system
during $F_1, \bar F_1$ production
era does not greatly modify
($\lesssim 20\%$)
our derived limits on $\epsilon$ from the constraints on $\delta
N_{\text{eff}}$.}
This means that $dS _{\nu} \simeq 0$. Nevertheless, the evolution of
$T _{\nu}$ will still have to be taken into account, though it trivially
scales as the inverse of the scale factor (see e.g.
\cite{earlyuniverse}).

The second law of thermodynamics states that the change in entropy in
the visible sector is given by:
\begin{eqnarray}
dS = -\frac{dQ}{T _{\gamma}} \ .
\label{entropyordinary}
\end{eqnarray}
Similarly, the change in entropy for the dark sector is:
\begin{eqnarray}
dS ^{'} = \frac{dQ}{T _{\gamma _{_D}}} \ .
\label{entropydark}
\end{eqnarray}
A useful way to express the entropy of a particle species in cosmology
is given in e.g. \cite{earlyuniverse}:
\begin{eqnarray}
S = \frac{\rho + p}{T}R^3 \ ,
\label{entropycosmology}
\end{eqnarray}
where $\rho$, $p$ and $T$ denote its energy density, pressure and
temperature respectively. Taking the derivative with respect to time on
both sides of Eqs.(\ref{entropyordinary},\ref{entropydark}) and
combining the result with Eqs.(\ref{dqdt},\ref{entropycosmology})
yields:
\begin{eqnarray}
\frac{d}{dt} \left [ \frac{(\rho _{\gamma} + p _{\gamma} + \rho _e + p
_e)R^3}{T _{\gamma}}\right ] 
&=& - \frac{n _e n _{\overline{e}} \langle \sigma v_{\text{M\o l}} {\cal
E} \rangle R^3}{T _{\gamma}} \ , \nonumber \\
\frac{d}{dt} \left [ \frac{(\rho _{\gamma _{_D}} + p _{\gamma _{_D}} +
\rho _{F_1} + p _{F_1})R^3}{T _{\gamma _{_D}}}\right ] 
&=& \frac{n _e n _{\overline{e}} \langle \sigma v_{\text{M\o l}} {\cal
E} \rangle R^3}{T _{\gamma _{_D}}} \ ,
\label{entropy1}
\end{eqnarray}
where we have neglected the neutrino contribution to the entropy change,
which is justified as $dS _{\nu} \simeq 0$, as discussed above. In
Eqs.(\ref{entropy1}) and below, we have defined $\rho _e \equiv \rho _e
+ \rho _{\bar{e}}$, and similarly for $p _e$, $\rho _{F_1}$ and $p
_{F_1}$.

The $^{00}$ component of the Einstein field equations for the FRW metric
describes the evolution of the scale factor $R$. This is known as first
Friedmann equation, and in a flat Universe takes the form:
\begin{eqnarray}
\left ( \frac{\dot{R}}{R} \right ) ^2 = \frac{8\pi G_N}{3} \left [\rho
_{\gamma} + \rho _e + \rho _{\nu} + \rho _{\gamma _{_D}} + \rho _{F_1}
\right ] \ .
\label{Friedmann}
\end{eqnarray}
Defining $x \equiv m _e/T _{\gamma}$, energy densities and pressures in
the visible sector are given in e.g. \cite{earlyuniverse}:
\begin{eqnarray}
\rho _{\gamma} &=& \frac{\pi ^2}{15} T _{\gamma} ^4 \ , \nonumber \\
p _{\gamma} &=& \frac{\pi ^2}{45} T _{\gamma} ^4 \ , \nonumber \\
\rho _e &=& \frac{2T _{\gamma} ^4}{\pi ^2} \mathlarger \int _{x}
^{\infty}du \ \frac{(u^2 - x^2)^{\frac{1}{2}}u^2}{1 + e^u} \ , \nonumber
\\
p _e &=& \frac{2T _{\gamma} ^4}{3\pi ^2} \mathlarger \int _{x}
^{\infty}du \ \frac{(u^2 - x^2)^{\frac{3}{2}}}{1 + e^u} \ , \nonumber \\
\rho _{\nu} &=& \frac{7\pi ^2}{40} T _{\nu} ^4 \ .
\end{eqnarray}
Similarly for the dark sector, with $x ^{'} \equiv m _{F_1}/T _{\gamma
_{_D}}$:
\begin{eqnarray}
\rho _{\gamma _{_D}} &=& \frac{\pi ^2}{15} {T _{\gamma _{_D}}}^4 \ ,
\nonumber \\
p _{\gamma _{_D}} &=& \frac{\pi ^2}{45} {T _{\gamma _{_D}}} ^4 \ ,
\nonumber \\
\rho _{F_1} &=& \frac{2{T _{\gamma _{_D}}} ^4}{\pi ^2} \mathlarger \int
_{x'} ^{\infty}du \ \frac{(u^2 - {x'}^2)^{\frac{1}{2}}u^2}{1 + e^u} \ ,
\nonumber \\
p _{F_1} &=& \frac{2{T _{\gamma _{_D}}} ^4}{3\pi ^2} \mathlarger \int
_{x'} ^{\infty}du \ \frac{(u^2 - {x'}^2)^{\frac{3}{2}}}{1 + e^u} \ .
\end{eqnarray}

Considering the neutrino subsystem, the neutrino temperature scales as
$T _{\nu} \propto 1/R$ which follows from $dS _{\nu} \simeq 0$. Noting
that all proportionality factors cancel [being there the same power of
the scale factor $R$ on both sides of Eqs.(\ref{entropy1})], $R$ in
Eqs.(\ref{entropy1}) can effectively be replaced by $1/T _{\nu}$.
Accordingly, Eqs.(\ref{entropy1}) can be expressed as:
\begin{eqnarray}
\frac{d}{dt} \left [ \frac{(\rho _{\gamma} + p _{\gamma} + \rho _e + p
_e)}{T _{\gamma} T _{\nu} ^3}\right ] 
&=& - \frac{n _e n _{\overline{e}} \langle \sigma v_{\text {M\o l}}
{\cal E} \rangle }{T _{\gamma} T _{\nu} ^3} \ , \nonumber \\
\frac{d}{dt} \left [ \frac{(\rho _{\gamma _{_D}} + p _{\gamma _{_D}} +
\rho _{F_1} + p _{F_1})}{T _{\gamma _{_D}}T _{\nu} ^3}\right ] 
&=& \frac{n _e n _{\overline{e}} \langle \sigma v_{\text{M\o l}} {\cal
E} \rangle }{T _{\gamma _{_D}}T _{\nu} ^3} \ ,
\label{system1}
\end{eqnarray}
and Eq.(\ref{Friedmann}) as:
\begin{eqnarray}
\frac{1}{T _{\nu}} \frac{dT _{\nu}}{dt} &=& - \sqrt{\frac{8\pi G_N}{3}
\left ( \rho _{\gamma} + 
\rho _e + \rho _{\nu} + \rho _{\gamma _{_D}} + \rho _{F_1} \right )} \ .
\label{friedmannbis}
\end{eqnarray}
Some manipulation shows that Eqs.(\ref{system1}) can be brought to the
form:
\begin{eqnarray}
\zeta \frac{dT _{\gamma}}{dt} + \kappa \frac{dT _{\nu}}{dt} 
&=& - \frac{n _{e} n _{\overline{e}} \langle \sigma v _{\text{M\o l}}
{\cal E} \rangle }{T _{\gamma} ^3} \ , \nonumber \\
\zeta ^{'} \frac{dT _{\gamma _{_D}}}{dt} + \kappa ^{'} \frac{dT
_{\nu}}{dt} 
&=& \frac{n _{e} n _{\overline{e}} \langle \sigma v _{\text{M\o l}}
{\cal E} \rangle }{T _{\gamma _{_D}} ^3} \ ,
\label{sistema1}
\end{eqnarray}
where:
\begin{eqnarray}
\zeta & \equiv &  \frac{3\rho _{\gamma}}{T _{\gamma} ^4} + \frac{3p
_{\gamma}}{T _{\gamma} ^4} + \frac{3\rho _e}{T _{\gamma} ^4} + \frac{3p
_e}{T _{\gamma} ^4} + \frac{2m _{e} ^2}{\pi ^2 T _{\gamma} ^2} \int _{x}
^{\infty} du \ \frac{(u^2 - x^2)^{-\frac{1}{2}} u^2 + (u^2 -
x^2)^{\frac{1}{2}}}{1 + e^u} \ , \nonumber \\
\kappa & \equiv & - \left ( \frac{3\rho _{\gamma}}{T _{\gamma} ^3} +
\frac{3p _{\gamma}}{T _{\gamma} ^3} + \frac{3\rho _e}{T _{\gamma} ^3} +
\frac{3p _e}{T _{\gamma} ^3} \right ) \frac{1}{T _{\nu}} \ , \nonumber
\\
\zeta ^{'} & \equiv & \frac{3\rho _{\gamma _{_D}}}{{T _{\gamma _{_D}}}
^4} + \frac{3p _{\gamma _{_D}}}{{T _{\gamma _{_D}}}^4} + \frac{3\rho
_{F_1}}{{T _{\gamma _{_D}}} ^4} + \frac{3p _{F_1}}{{T _{\gamma _{_D}}}
^4} + \frac{2m _{F_1} ^2}{\pi ^2 {T _{\gamma _{_D}}} ^2} \int _{x'}
^{\infty} du \ \frac{(u^2 - {x'}^2)^{-\frac{1}{2}} u^2 + (u^2 -
{x'}^2)^{\frac{1}{2}}}{1 + e^u} \ , \nonumber \\
\kappa ^{'} & \equiv & - \left ( \frac{3\rho _{\gamma _{_D}}}{{T
_{\gamma _{_D}}}^3} + \frac{3p _{\gamma _{_D}}}{{T _{\gamma _{_D}}} ^3}
+ \frac{3\rho _{F_1}}{{T _{\gamma _{_D}}} ^3} + \frac{3p _{F_1}}{{T
_{\gamma _{_D}}} ^3} \right ) \frac{1}{T _{\nu}} \ .
\label{sistemo}
\end{eqnarray}

\begin{figure}[htpb]
\vskip 1.0cm
    \centering
        \includegraphics[scale=0.5, angle=270]{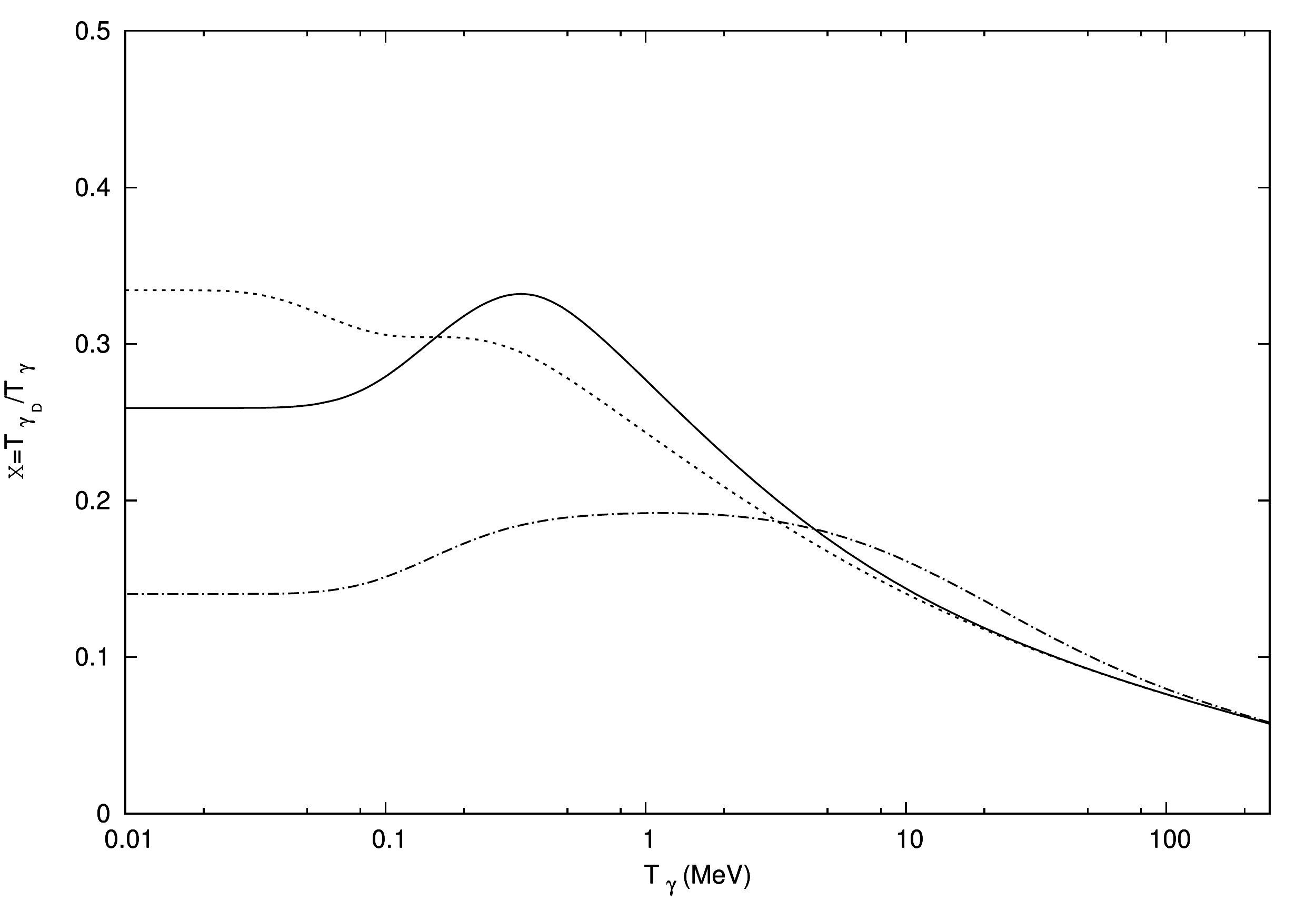}
    \caption{Evolution of ${\cal X} \equiv T _{\gamma _{_D}}/T
_{\gamma}$ for $m_{F_1}=10$ MeV (dot-dashed line), $m_{F_1}=1$ MeV
(solid line) and $m_{F_1}=0.1$ MeV (dashed line).}
    \label{fig:evolution of x}
\vskip 0.9cm
\end{figure}

We are now left with a closed system of three differential equations
[Eqs.(\ref{friedmannbis},\ref{sistema1})] for three unknowns ($T
_{\gamma}$, $T _{\gamma _{_D}}$ and $T _{\nu}$). Given suitable initial
conditions, then, the system can be solved numerically to give the
evolution of these three quantities. An example is presented in Figure
\ref{fig:evolution of x}, where the evolution of $T _{\gamma _{_D}}/T
_{\gamma}$ is plotted as a function of $T _{\gamma}$ for different
values of $m _{F_1}$ and for $\epsilon = 10 ^{-9}$. Note that the flow
of time is from the right to the left.

It can be seen from Figure \ref{fig:evolution of x} that $T _{\gamma
_{_D}}/T _{\gamma}$ asymptotically approaches a constant at late times.
We would like to find an approximate analytic expression for the
asymptotic value of $T _{\gamma _{_D}}/T _{\gamma}$. It is perhaps
useful to recall the results obtained for MDM. In this context, MDM can
be viewed as a special case of our model in the limit where $m _{F_1} =
m _e$. For the case of MDM it has been found that $T _{{\gamma} ^{'}}/T
_{\gamma}$ (where $\gamma ^{'}$ denotes the mirror photon, which is of
course analogous to our dark photon, $\gamma _{_D}$) asymptotically
evolves to \cite{predictions}:
\begin{eqnarray}
\frac{T _{{\gamma} ^{'}}}{T _{\gamma}} \simeq 0.31 \left (
\frac{\epsilon}{10^{-9}} \right ) ^{\frac{1}{2}} \ .
\label{xmirror}
\end{eqnarray}
More generally, $m _{F_1} \neq m _e$ in the context of our two-component
hidden sector model, and one expects a somewhat different behavior in $T
_{\gamma _{_D}}/T _{\gamma}$ to account for this $m _{F_1}$ dependence.
Previous work in the MDM context shows that in the limit of $T _{\gamma}
\gg m _e$, an analytic expression can be found for $T _{{\gamma} ^{'}}/T
_{\gamma}$ \cite{ciarcellutiliege}:
\begin{eqnarray}
\frac{T _{{\gamma} ^{'}}}{T _{\gamma}} \propto \sqrt{\epsilon} \left [
\frac{1}{T} - \frac{1}{T_i} \right ] ^{\frac{1}{4}} \ ,
\label{epsilont}
\end{eqnarray}
with an assumed initial condition $T _{{\gamma} ^{'}} = 0$ at $T
_{\gamma} = T_i$. For $T _{\gamma} \sim m _e$, energy transfer to the
mirror sector cuts off, as the process $e\overline{e} \rightarrow e
^{'}\overline{e} ^{'}$ becomes infrequent due to Boltzmann suppression
of $e$, $\overline{e}$ number densities.

We can attempt to generalize the result to our case. The process
$e\overline{e} \rightarrow F _1\overline{F} _1$ will cease to be
important at temperatures below $\sim {\cal M} \equiv \max (m _e , m
_{F_1})$. Eq.(\ref{epsilont}) then suggests that the asymptotic value of
the ratio $T _{\gamma _{_D}}/T _{\gamma}$ is proportional to
$\sqrt{\epsilon} \left (m _e/{\cal M} \right ) ^{\frac{1}{4}}$. This
intuition has been verified numerically, by evolving for different
values of $\epsilon$ and $m _{F_1}$. Numerically, we find that the
asymptotic value of $T _{\gamma _{_D}}/T _{\gamma}$ can be expressed in
the form:
\begin{eqnarray}
\frac{T _{\gamma _{_D}}}{T _{\gamma}} & \simeq & 0.31
\sqrt{\frac{\epsilon}{10^{-9}}} \left (\frac{m _e}{{\cal M}} \right )
^{\frac{1}{4}} \ , \nonumber \\
{\cal M} & \equiv & \max (m _e, m _{F_1}) \ ,
\label{masterformula}
\end{eqnarray}
for parameters in the range $\epsilon \sim 10 ^{-9}$ and $0.1$ MeV
$\lesssim m _{F_1} \lesssim 100$ MeV.

One can also attempt to understand the shape of the curves in Figure
\ref{fig:evolution of x}. At early times ($T _{\gamma} \gg m _{F_1},m
_e$) the curves overlap, following a $T _{\gamma _{_D}}/T _{\gamma}
\propto \left (1/T _{\gamma} \right ) ^{\frac{1}{4}}$ behavior
consistent with the analytic solution previously discussed. At some
later time corresponding to $T _{\gamma} \sim {\cal M}$, the curves
start deviating from the analytic solution. The rising of the various
curves at different temperatures and with characteristic bumps can be
understood in terms of annihilation processes which are heating the
respective sectors roughly at the temperature corresponding to the mass
of the particle-antiparticle pair which is annihilating. That is,
electron-positron and $F _1$-$\overline{F} _1$ annihilations explain the
deviation of the numerical solution from the simpler analytic one. Once
the annihilation processes are over, $T _{\gamma _{_D}}/T _{\gamma}$
reaches its asymptotic value.

\vskip 3.4cm
\subsection{Calculation of $\delta N _{\text{eff}}$[CMB]}
\vskip 0.2cm

We now compute the modification of the energy density at the Hydrogen
recombination epoch in the early Universe. A way to parameterize this
extra energy density is in terms of an effective number of neutrino
species. Recall that the relativistic energy density component at
recombination can be expressed as:
\begin{eqnarray}
\rho _{\text{rad}} = \left (1 + \frac{7}{8} \left (\frac{4}{11} \right)
^{\frac{4}{3}} N _{\text{eff}}[\text{CMB}] \right ) \rho _{\gamma} \ ,
\end{eqnarray}
where the factor of $7/8$ takes into account the different statistical
nature (fermionic instead of bosonic) of neutrinos with respect to
photons, and the
factor of $4/11$ takes care of $\gamma$ heating due to $e \overline{e}$
annihilation after neutrino kinetic decoupling (see for instance
\cite{earlyuniverse}). $N _{\text{eff}}$ is referred to as the effective
number of neutrinos, and is predicted to be $N _{\text{eff}} \simeq
3.046$ in the
Standard Model (see e.g. \cite{mangano}). 
Observations from WMAP \cite{wmapx}, the South Pole Telescope
\cite{sptx}, the Atacama Cosmology Telescope
\cite{atax} and the Planck mission \cite{plax} are consistent with the
Standard Model predictions and can be used to constrain 
$\delta N_{\text{eff}}[\text{CMB}] \equiv N _{\text{eff}}[\text{CMB}] -
3.046$. 
Using the result of 
Planck's analysis $N_{\text{eff}}$[CMB] = $3.30 \pm 0.27$ \cite{plax},
gives the  
$2\sigma$ upper limit: $\delta N _{\text{eff}}[\text{CMB}] < 0.80$. 

In our model the modification to the effective number of neutrinos can
be written as follows:
\begin{eqnarray}
\delta N _{\text{eff}}[\text{CMB}] = 3 \left (\left [\frac{T _{\nu} (
\epsilon )}{T _{\nu} ( \epsilon = 0)}\right ]^4 - 1 \right ) +
\frac{8}{7} \left ( \frac{T _{\gamma _{_D}} ( \epsilon )}{T _{\nu} (
\epsilon = 0)} \right )^4 \ ,
\label{Neffcmb}
\end{eqnarray}
where the temperatures are evaluated at photon decoupling, $T _{\gamma}
\simeq 0.26$ eV. The two terms on the right-hand side of
Eq.(\ref{Neffcmb}) account for distinct effects. Firstly, the process $e
\overline{e} \rightarrow F_1 \overline{F}_1$ will increase $T _{\gamma
_{_D}}$ at the expense of $T _{\gamma}$, thus reducing $T _{\gamma}/T
_{\nu}$ and effectively increasing the number of neutrino species at
recombination. The second term is the direct increase in $N
_{\text{eff}}[\text{CMB}]$ due to the increase in $T _{\gamma _{_D}}$
itself.

One has to pay attention when using $\delta N _{\text{eff}}$[CMB] to set
constraints on the parameter space, since the addition of energy density
is not the
only effect to consider. Prior to recombination of $F _1$ and $F _2$
into neutral dark states, dark matter behaves like a tightly coupled
fluid, analogous to
the photon-baryon fluid in the visible sector. This fluid undergoes
acoustic oscillations, which suppress power on small scales, hence
behaving very
differently from collisionless CDM. Thus, there are two quite different
possible effects for the CMB to consider. The first is the extra energy
density as
parameterized by $\delta N _{\text{eff}}$[CMB], and the second is the
effect of dark acoustic oscillations prior to dark recombination. In
this section we consider the energy density modification, while the
constraints arising from dark acoustic oscillations will be dealt with
in Section 4.

\begin{figure}[htpb]
    \centering
        \includegraphics[scale=0.5, angle=270]{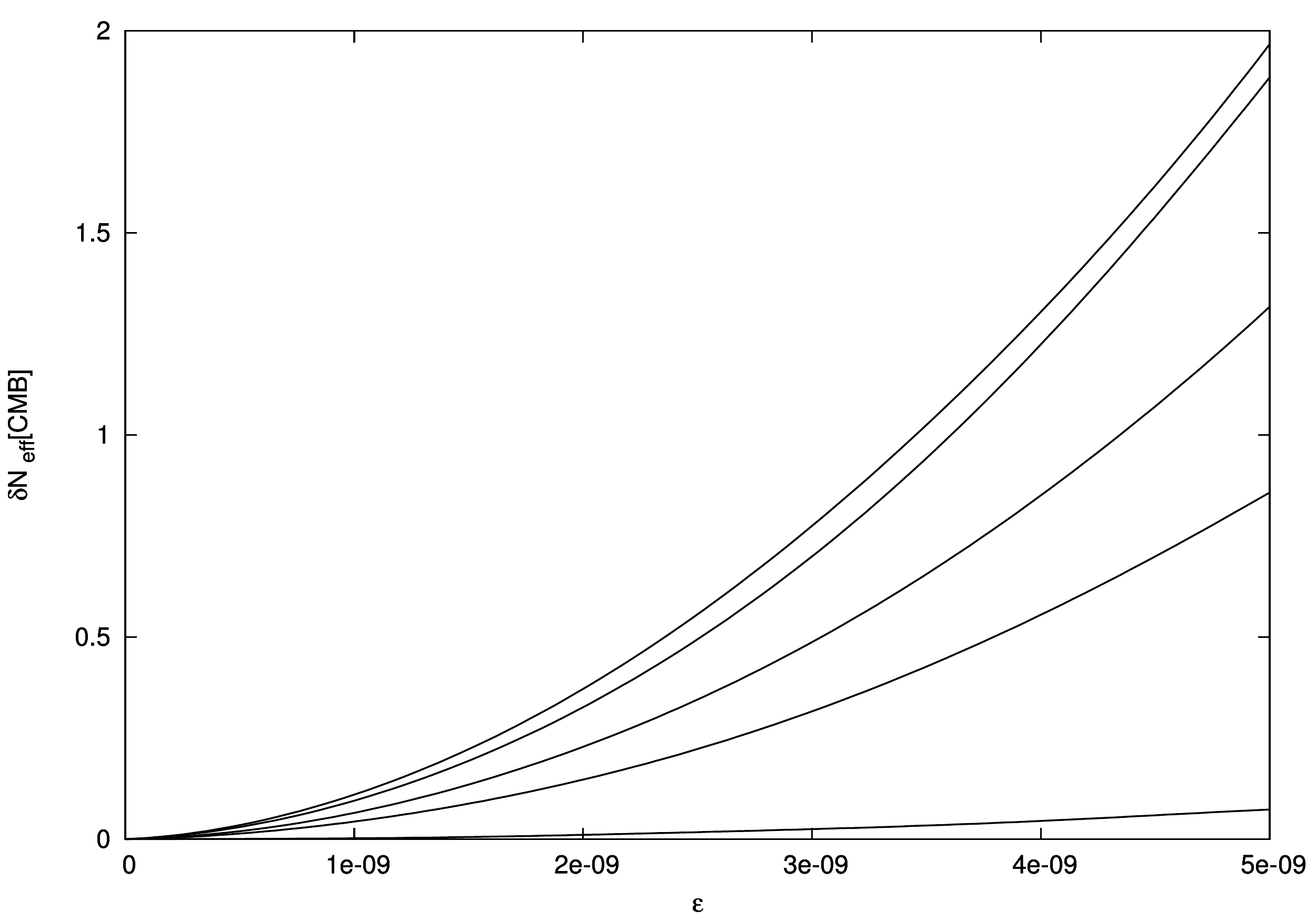}
    \caption{$\delta N _{\text{eff}}$[CMB] versus $\epsilon$ at fixed
values of $m_{F_1}$ for (going from up to down) $m_{F_1}=0.1, 0.511,
0.7, 1, 10$ MeV.}
    \label{fig:Neffcmb}
\end{figure}
\begin{figure}[htpb]
\vskip -1.2cm
    \centering
        \includegraphics[scale=0.5, angle=270]{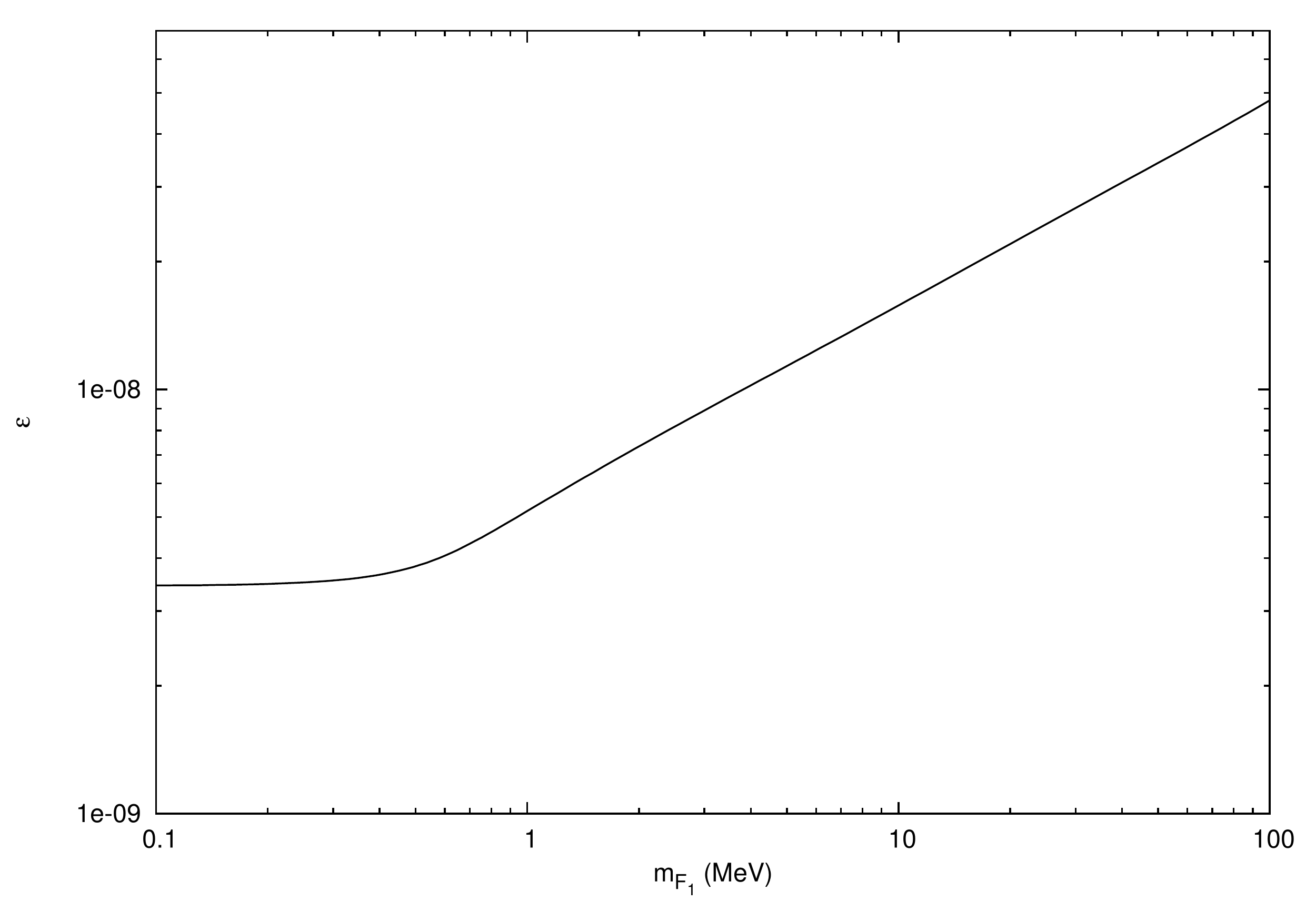}
    \caption{Exclusion limits obtained from $\delta N
_{\text{eff}}$[CMB] $<$ 0.80 in $\epsilon$-$m_{F_1}$ parameter space
(excluded region is above line).}
    \label{fig:Exclusion cmb}
\end{figure}

In Figure \ref{fig:Neffcmb}, we present results for $\delta N
_{\text{eff}}$[CMB] obtained by numerically solving 
Eq.(\ref{Neffcmb}) [in the process, solving also
Eqs.(\ref{friedmannbis},\ref{sistema1})] for some example parameter
choices. We set constraints on our model by using the limit $\delta N
_{\text{eff}}[\text{CMB}] < 0.80$. In Figure \ref{fig:Exclusion cmb} the
exclusion limits for our model in the $\epsilon$-$m _{F_1}$ parameter
space are shown, with the excluded region being above the line. Notice
for $m _{F_1} = 0.511$ MeV we recover the bound on $\epsilon$ obtained
for MDM, $\epsilon \lesssim 3.5 \times 10 ^{-9}$ \cite{predictions}.

%
%
%
\vskip 3cm
\subsection{Calculation of $\delta N _{\text{eff}}$[BBN]}

The addition of extra energy density during the early Universe also has
an effect on BBN, the process during which light nuclei, and in
particular helium, were synthesized (for a more detailed review see e.g.
\cite{weinberg}). It is known that increasing the energy density by the
addition of one neutrino species increases the helium fraction, $Y _p$,
by approximately 0.013 \cite{bernstein}. It follows therefore that the
change in the effective number of neutrino species associated with BBN
is approximately given by:
\begin{eqnarray}
\delta N _{\text{eff}}[\text{BBN}] = \frac{Y _p (\epsilon) - Y _p
(\epsilon = 0)}{0.013} \ .
\label{Neffbbn}
\end{eqnarray}
The first step towards the synthesis of helium is the synthesis of
deuterium which, in turn, depends on the neutron abundance $X _n \equiv
n _p/(n _n + n _p)$. We begin by considering the weak interaction
processes which affect the neutron abundance:
\begin{eqnarray}
n + \nu _e \leftrightarrow p + e \ , \nonumber \\
n + \bar e \leftrightarrow p + \bar \nu _e \ , \nonumber \\
n \rightarrow p + e + \bar \nu _e \ .
\end{eqnarray}
At equilibrium (hence, at high temperatures) $X _n \simeq 1/(1 +
e^{Q/T})$, where $Q \simeq 1.293$ MeV is the difference between the
neutron and the proton mass.

The rates for the four processes which affect the neutron abundance
(excluding neutron decay) can be found in e.g. \cite{weinberg}:
\begin{eqnarray}
\lambda _1 & \equiv & \lambda (n + \nu _e \rightarrow p + e) = A
\mathlarger \int ^{\infty}_{0} dP_{\nu} \ E_e^2 P_{\nu}^2
\frac{1}{e^{\frac{E_{\nu}}{T_{\nu}}}+1}
\frac{1}{e^{-\frac{E_e}{T_{\gamma}}}+1} \ , \nonumber
\\
\lambda _2 & \equiv & \lambda (n + \bar e \rightarrow p + \bar \nu _e) =
A \mathlarger \int ^{\infty}_{0} dP_e \ E_{\nu}^2 P_e^2
\frac{1}{e^{\frac{E_e}{T_{\gamma}}}+1}
\frac{1}{e^{-\frac{E_{\nu}}{T_{\nu}}}+1} \ , \nonumber
\\
\lambda _3 & \equiv & \lambda (p + e \rightarrow n + \nu _e) = A
\mathlarger \int ^{\infty}_{\sqrt{Q^2-m_e^2}} dP_e \ E_{\nu}^2 P_e^2
\frac{1}{e^{\frac{E_e}{T_{\gamma}}}+1}
\frac{1}{e^{-\frac{E_{\nu}}{T_{\nu}}}+1} \ , \nonumber
\\
\lambda _4 & \equiv & \lambda (p + \bar \nu _e \rightarrow n + \bar e) =
A \mathlarger \int ^{\infty}_{Q+m_e} dP_{\nu} \ E_e^2 P_{\nu}^2
\frac{1}{e^{\frac{E_{\nu}}{T_{\nu}}}+1}
\frac{1}{e^{-\frac{E_e}{T_{\gamma}}}+1} \ ,
\end{eqnarray}
where $E _e$[$E _{\nu}$], $P _e$[$P _{\nu}$] indicate the electron
[neutrino] energy and momentum respectively. The extremals of the
integrals are obtained from kinematical considerations. The factors
within the integrals account for Fermi-Dirac statistics and Pauli
blocking. The values of the various constants are given by:
\begin{eqnarray}
A & = & \frac{G_F^2 (1+3g_A^2)\cos ^2\theta _c}{2\pi ^3} \ , \nonumber
\\
G_F & = & 1.166 \times 10^{-5} \ {\rm GeV^{-2}} \ , \nonumber \\
g_A & = & 1.257 \ , \nonumber \\
\cos \theta _c & = & 0.97456 \ .
\end{eqnarray}
The evolution of the neutron abundance, $X _n$, is governed by the differential equation:
\begin{eqnarray}
\frac{dX _n}{dt} = -( \lambda _1 + \lambda _2 + \lambda _n)X _n +
(\lambda _3 + \lambda _4)(1 - X _n) \ ,
\label{neutron}
\end{eqnarray}
where $\lambda _n ^{-1} = \tau _n \simeq 886.7 \ \rm s$ is the neutron
lifetime. Eq.(\ref{neutron}) can be used to evolve the neutron fraction
down to the so-called \textit{deuterium bottleneck} temperature $T
_{\gamma} \simeq$ 0.07 MeV (of course,
Eqs.(\ref{friedmannbis},\ref{sistema1}) need to be solved simultaneously
to obtain the modified time-temperature relation). The helium fraction,
$Y _p$, is twice the value of $X _n$ at this time, and $\delta N
_{\text{eff}}$[BBN] can be evaluated by using Eq.(\ref{Neffbbn}).

\begin{figure}[htpb]
    \centering
        \includegraphics[scale=0.5, angle=270]{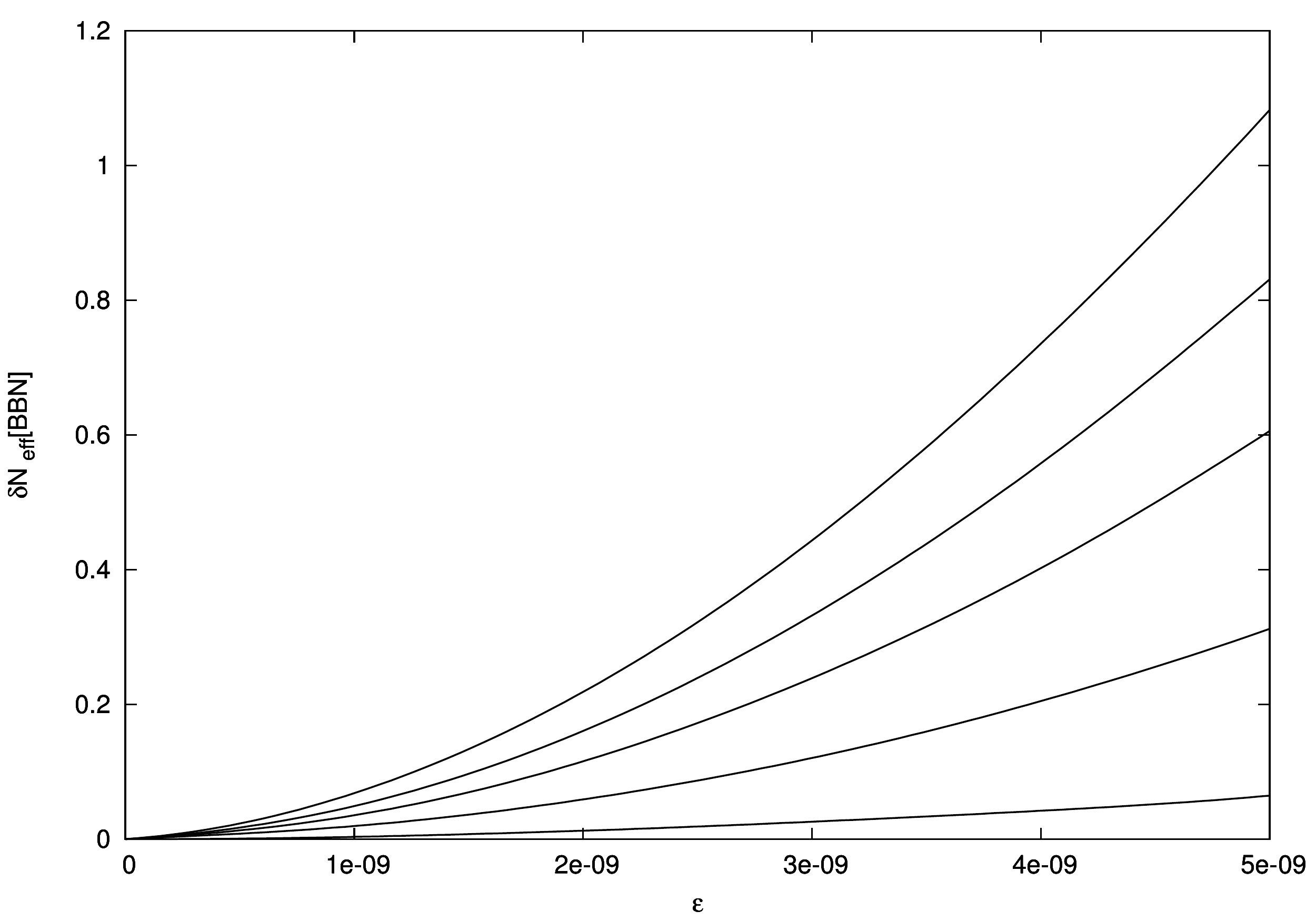}
    \caption{$\delta N _{\text{eff}}$[BBN] versus $\epsilon$ at fixed
values of $m_{F_1}$ for (going from up to down) $m _{F_1} = 0.1, 0.7, 1,
2, 10$ MeV.}
    \label{fig:Neffbbnmf1}
\end{figure}

\begin{figure}[htpb]
    \centering
        \includegraphics[scale=0.5, angle=270]{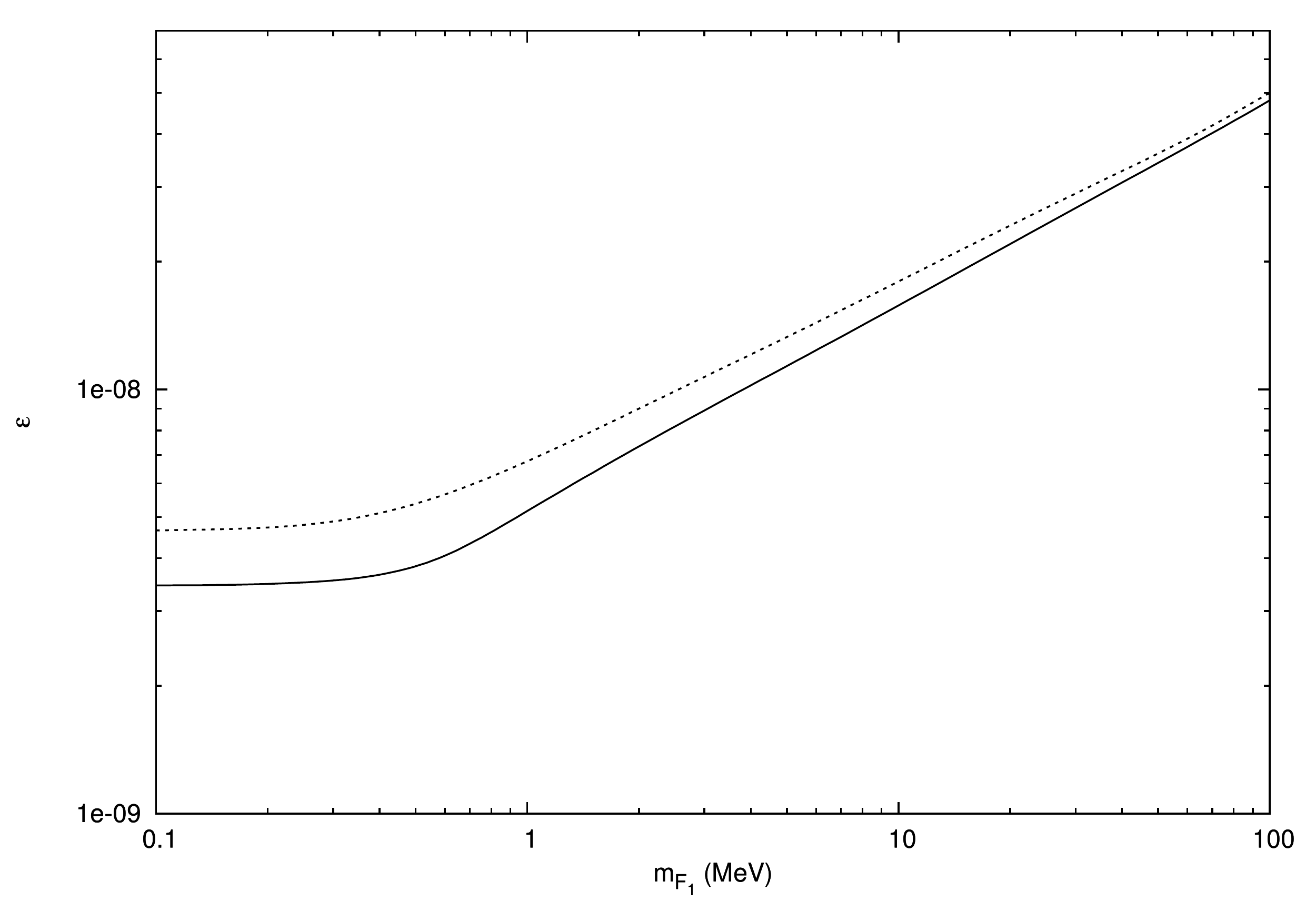}
    \caption{Exclusion limits from $\delta N _{\text{eff}}$[CMB] $<$
0.80 (solid line) and $\delta N _{\text{eff}}$[BBN] $<$ 1 (dashed line).
Region above lines are excluded.}
    \label{fig:Comparison bbn cmb}
\end{figure}

There are hints that $\delta N _{\text{eff}}$[BBN] is also non-zero and
positive. The data constrains $\delta N _{\text{eff}}[\text{BBN}] < 1$
at around 95\% confidence level \cite{izotov}. In Figure
\ref{fig:Neffbbnmf1} $\delta N _{\text{eff}}$[BBN] is plotted against
$\epsilon$ keeping $m _{F_1}$ fixed. The constraints following from this
analysis are shown together with those obtained from $\delta N
_{\text{eff}}$[CMB] in Figure \ref{fig:Comparison bbn cmb}. Evidently
the limits set by $\delta N _{\text{eff}}$[CMB] are more stringent than
those set by $\delta N _{\text{eff}}$[BBN]. Finally, we find an analytic
approximation to CMB $\delta N _{\text{eff}}$ constraints on $\epsilon$
arising from early Universe cosmology:
\begin{eqnarray}
\epsilon \lesssim 3.5 \times 10 ^{-9} \left ( \frac{{\cal M}}{m _e}
\right ) ^{\frac{1}{2}} \ .
\label{constraints}
\end{eqnarray}
The $\epsilon \sim {\cal M}^{\frac{1}{2}}$ dependence can easily be
understood by referring to Eqs.(\ref{masterformula},\ref{Neffcmb}).

\vskip 1 cm

\newpage

\section{Dark recombination}

\subsection{Saha equation for dark recombination}

Additional energy density, as parameterized by $\delta N
_{\text{eff}}$[CMB], is not the only new physics affecting the CMB.
Prior to recombination of $F _1$ and $F _2$ into neutral dark states,
dark matter behaves like a tightly coupled fluid which undergoes dark
acoustic oscillations. These oscillations suppress power on small scales
(hence deviating from collisionless CDM), below some characteristic
scale $L ^{\star}$, which is itself a function of the parameters in our
model. Ultimately such a suppression of power on small scales may help
in explaining the observed dearth of small galaxies in the neighborhood
of the Milky Way. In the following, though, we simply derive approximate
bounds by requiring that $T _{\text{dr}} \gtrsim T _{\text{eq}}$, where
$T _{\text{dr}}$ is the temperature in the visible sector at the time of
\textit{dark recombination}, and $T _{\text{eq}}$ is the temperature of
matter-radiation equality. This requirement has been used in the
literature (see for instance \cite{volkaspetraki}), and follows from
studies in the MDM context \cite{berezhiani,implications,cia}. Roughly,
$T _{\text{dr}} \gtrsim T _{\text{eq}}$ means that LSS is unaffected by
dark acoustic oscillations on scales which are still growing linearly
today.

In the present model, \textit{dark recombination} involves $|Z ^{'}|$ $F
_1$ particles combining with one $F _2$ particle to form a $U(1)
^{'}$-neutral dark state, which will be called $D ^0$ (recall $Z ^{'}$
is the charge ratio of $F _2$ and $F _1$). We would like to know when
(at what temperature or, equivalently, redshift) does dark recombination
happen, that is, the moment in which the \textit{last} $F _1$ recombines
with the state formed by $|Z ^{'}|$-1 $F _1$ particles and one $F _2$.
Let us call this last state $D ^+$ (we take the convention where $F _1$
has charge -1 and $F _2$ has charge $|Z'|$). The relevant process to
look at is:
\begin{eqnarray}
F _1 + D ^+ \leftrightarrow D ^0 + \gamma _{_D} \ .
\end{eqnarray}
The Saha equation for the process above is given in e.g.
\cite{dodelson}:
\begin{eqnarray}
\frac{n _{D ^0}}{n _{D ^+} n _{F_1}} = \frac{{n _{D ^0}}^{(0)}}{{n_{D
^+}}^{(0)}{n _{F_1}}^{(0)}} \ ,
\label{Saha}
\end{eqnarray}
where the superscript $^{(0)}$ denotes the equilibrium value. Note that
in writing Eq.(\ref{Saha}) it has been assumed $n _{\gamma _{_D}} = {n
_{\gamma _{_D}}} ^{(0)}$. It is worth stressing that Eq.(\ref{Saha}) is
an approximate equilibrium equation, namely, the equilibrium limit of
the Boltzmann equations. It does not, therefore, follow the abundances
through out-of-equilibrium processes, such as freeze-out (see for
instance \cite{dodelson}). Eq.(\ref{Saha}), nonetheless, predicts the
correct redshift of dark recombination, which is the quantity we wish to
determine.

To proceed, it is useful to introduce the ionization fraction of $F _1$:
\begin{eqnarray}
\chi \equiv \frac{n _{F_1}}{n _{F_2}} = \frac{n _{F_1}}{n _{F_1} + n _{D
^0}} = \frac{n _{F_1}}{n _{D ^+} + n _{D ^0}} \ ,
\end{eqnarray}
where $n _{F_1}$ is the number density of free $F_1$ particles and $n
_{F_2}$ is the total number density of $F_2$. The last equality follows
from assuming $U(1) ^{'}$ neutrality. The left-hand side of
Eq.(\ref{Saha}) is then $(1-\chi)/(n _{F_2}\chi ^2)$. The right-hand
side of Eq.(\ref{Saha}) can also be expressed in a more useful form. For
a species $A$ of mass $m _A$ and temperature $T _A$, the equilibrium
number density in the limit $m _A \gg T _A$ can be written as (see e.g.
\cite{dodelson}):
\begin{eqnarray}
n _A = g _A \left ( \frac{m _A T _A}{2\pi} \right ) ^{\frac{3}{2}} e
^{-\frac{m _A - \mu _A}{T _A}} \ ,
\end{eqnarray}
with $\mu _A$ being the chemical potential of the species and $g _A$ a
degeneracy factor that usually takes into account multiple spin states.
To good approximation $\mu _{\gamma _{_D}} = 0$ so, as long as
equilibrium holds, the following is true:
\begin{eqnarray}
\mu _{F_1} + \mu _{D ^+} = \mu _{D ^0} \ .
\end{eqnarray}
The ionization energy of $D ^0$, $I ^{'}$, is defined to be:
\begin{eqnarray}
I ^{'} = m _{F_1} + m _{D ^+} - m _{D ^0} \ .
\end{eqnarray}
Eq.(\ref{Saha}) can be rearranged in a form which is more useful for
following the evolution of the ionization fraction of $F _1$. To do so,
we can employ the fact that $g _{F_1}g _{D ^+} = g _{D ^0}$ and work in
the approximation $m _{D ^+} \simeq m _{D ^0} \simeq m _{F_2}$. This
approximation is valid as long as $m _{F_2} \gg m _{F_1}$ which is
assumed.\footnote{This approximation is similar to that of approximating
the mass of the Hydrogen atom with the proton mass.} These
considerations allow the right-hand side of Eq.(\ref{Saha}) to be
rearranged to the form:
\begin{eqnarray}
\frac{{n _{D ^0}}^{(0)}}{{n_{D ^+}}^{(0)}{n _{F_1}}^{(0)}} = \left (
\frac{2\pi}{m _{F_1}T _{\gamma _{_D}}} \right ) ^{\frac{3}{2}} e
^{\frac{I ^{'}}{T _{\gamma _{_D}}}} \ .
\label{Saha1}
\end{eqnarray}
The end result is that the Saha equation [Eq.(\ref{Saha})] can be
reduced to the more suitable form:
\begin{eqnarray}
\frac{1 - \chi}{\chi ^2} = n _{F_2}\left ( \frac{2\pi}{m _{F_1}T
_{\gamma _{_D}}} \right ) ^{\frac{3}{2}} e ^{\frac{I ^{'}}{T _{\gamma
_{_D}}}} \ .
\label{Saha2}
\end{eqnarray}
The $F _2$ number density simply scales with the baryon number density
as follows:
\begin{eqnarray}
n _{F_2} = \left ( \frac{\Omega _{\text{dm}}}{\Omega _{\text b}} \right
) \left ( \frac{m _p}{m _{F_2}} \right ) n _B = \left ( \frac{\Omega
_{\text{dm}}}{\Omega _{\text b}} \right ) \left ( \frac{m _p}{m _{F_2}}
\right ) \left ( \frac{n _B}{n _{\gamma}} \right ) \left ( \frac{n
_{\gamma}}{n _{\gamma _{_D}}} \right ) n _{\gamma _{_D}} \ ,
\label{nf2scale}
\end{eqnarray}
where $m _p \simeq 0.94 \ {\rm GeV}$ is the proton mass and $\eta \equiv
n _B/n _{\gamma}$ is the baryon-to-photon ratio. Using $\Omega
_{\text{dm}}/\Omega _{\text b} \simeq 5.4$ \cite{plax}, $\eta \simeq 6
\times 10 ^{-10}$ \cite{pdg}, $n _{\gamma}/n _{\gamma _{_D}} = \left (T
_{\gamma}/T _{\gamma _{_D}} \right ) ^3$ [with $T _{\gamma}/T _{\gamma
_{_D}}$ evaluated using Eq.(\ref{masterformula})] and $n _{\gamma _{_D}}
= \pi ^2T _{\gamma _{_D}} ^3/45$ allows us to rewrite Eq.(\ref{Saha2})
in the following form:
\begin{eqnarray}
\frac{1 - \chi}{\chi ^2} = A \left ( \frac{T _{\gamma _{_D}}}{I ^{'}}
\right ) ^{\frac{3}{2}} e ^{\frac{I ^{'}}{T _{\gamma _{_D}}}} \ ,
\label{Saha23}
\end{eqnarray}
where:
\begin{eqnarray}
A \simeq 3.5 \times 10 ^{-7} \left ( \frac{10 ^{-9}}{\epsilon} \right )
^{\frac{3}{2}} \left ( \frac{{\cal M}}{m _e} \right ) ^{\frac{3}{4}}
\left ( \frac{\rm GeV}{m _{F_2}} \right ) \left ( \frac{I ^{'}}{m
_{F_1}} \right ) ^{\frac{3}{2}} \ .
\end{eqnarray}
Using the variable $\xi \equiv I ^{'}/T _{\gamma _{_D}}$,
Eq.(\ref{Saha23}) can be put to the form:
\begin{eqnarray}
\frac{1 - \chi}{\chi ^2} = A \xi ^{-\frac{3}{2}} e ^{\xi} \ .
\label{Saha3}
\end{eqnarray}
The Saha equation can be used to determine the redshift of dark
recombination. To solve for the redshift (or, equivalently, temperature)
of dark recombination, we take $\chi \approx 0.1$, so that
Eq.(\ref{Saha3}) reduces to:
\begin{eqnarray}
\xi = \frac{3}{2} \ln \xi + \ln \left ( \frac{90}{A} \right ) \ .
\label{sahafinal}
\end{eqnarray}
In this form the Saha equation is easy to solve numerically. Once  the
value of $\xi$ that solves the equation has been found, the temperature
of the dark sector at dark recombination, $T _{\text{dr}} ^{'}$, is
given by:
\begin{eqnarray}
T _{\text{dr}} ^{'} = \frac{I ^{'}}{\xi} \ .
\label{tdark}
\end{eqnarray}
The corresponding temperature of the visible sector at the time of dark
recombination, $T _{\text{dr}}$, can be found by inverting
Eq.(\ref{masterformula}):
\begin{eqnarray}
T _{\text{dr}} \simeq 3.2 \ T _{\text{dr}} ^{'} \left ( \frac{10
^{-9}}{\epsilon} \right ) ^{\frac{1}{2}} \left ( \frac{{\cal M}}{m _e}
\right ) ^{\frac{1}{4}} \ .
\label{darkord}
\end{eqnarray}
\subsection{Binding energy of the dark bound state}

To make progress, we need to determine $I ^{'}$ in terms of the
parameters of our model. The bound system of $F _2$ with $N$ $F _1$
particles is completely analogous to that of nuclei with $N$ electrons.
It follows that the binding energy of the dark state has the general
form:
\begin{eqnarray}
I ^{'} = {Z _{\text{eff}}'} ^2 \frac{\alpha '^{2}}{2} \mu _R \ ,
\label{ip}
\end{eqnarray} 
where $\mu _R$ is the reduced mass of the $F_1$-$D ^+$ system, given by
$\mu _R = m _{F_1}m _{D ^+}/(m _{F_1} + m _{D ^+})$. In the limit where
$m _{F_2} \gg m _{F_1}$ one has that $I ^{'} \simeq Z _{\text{eff}}
^{'2} \ {\alpha '} ^2 m _{F_1}/2$.

Naturally exact analytic expressions for $Z _{\text{eff}} ^{'}$ are in
general unknown, but it is still possible to make a rough approximation
for $Z ^{'} _{\text{eff}}$ and hence determine $I ^{'}$. The charge $Z
_{\text{eff}} ^{'}$ depends only on the chemistry (or equivalently on
quantum mechanics) of the bound state we are analyzing. In particular,
it depends on shielding effects due to the $|Z ^{'}|$-$1$ $F _1$
particles partially shielding the charge of the $F _2$ particle from the
last $F _1$ which is about to combine. The problem of determining $Z
_{\text{eff}} ^{'}$ is therefore identical to that of determining the
shielding of an ordinary nucleus of atomic number $Z = |Z ^{'}|$ due to
$Z$-$1$ electrons. It essentially only depends on the way the fermions
arrange themselves in orbitals, which in turn is determined solely by
quantum mechanics.

\begin{figure}[htpb]
    \centering
        \includegraphics[scale=0.5, angle=270]{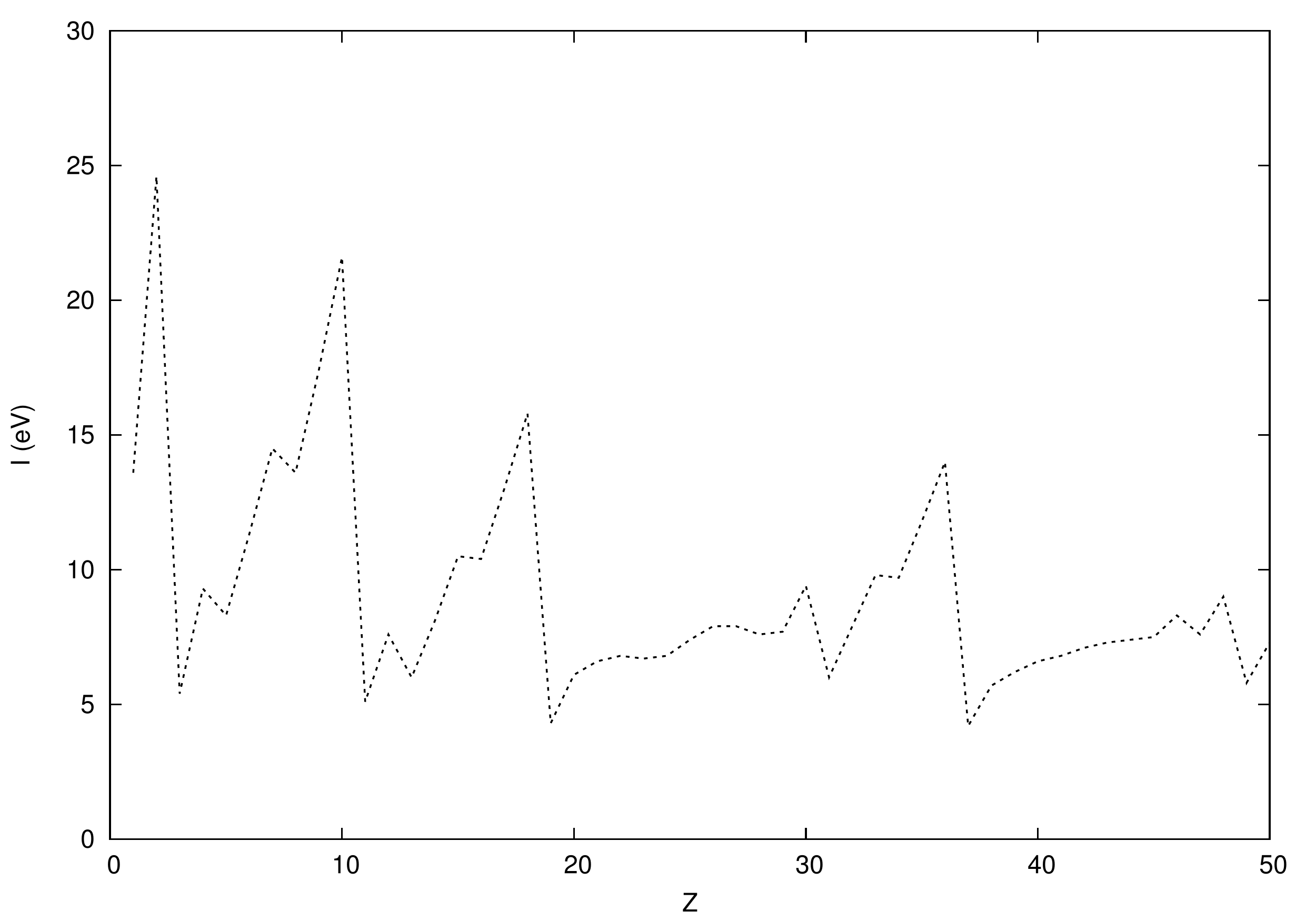}
    \caption{Ionization energy as a function of atomic number for
ordinary elements.}
    \label{fig:Ionization}
\end{figure}

Under these assumptions the binding energy $I ^{'}$ of the dark bound
state can be derived simply by scaling the binding energy $I$ of the
corresponding ordinary element with atomic number $Z = |Z ^{'}|$ via:
\begin{eqnarray}
I ^{'} = \left ( \frac{\alpha ^{'}}{\alpha} \right ) ^2 \left ( \frac{m
_{F_1}}{m _e} \right ) I \ .
\label{scaling}
\end{eqnarray}
A plot of the binding energies of the elements of the periodic table as
a function of the atomic number $Z$ is shown in Figure
\ref{fig:Ionization}. One notes from Figure \ref{fig:Ionization} that,
apart from isolated cases such as He, the binding energies of the
various elements reside in a fairly narrow range centered at about $10$
eV, within a factor of approximately 2. For $Z \gtrsim 10$, the
dependence of $I$ on $Z$ is even weaker. This means that $Z
_{\text{eff}} ^{'} \approx 1$ in Eq.(\ref{ip}) and $I ^{'} \approx
{\alpha '} ^2m _{F_1}/2$.

\vskip 1 cm

\newpage

\subsection{Exclusion limits}
\vskip 0.3cm

Recall the validity of our model requires $T _{\text{dr}} \gtrsim T
_{\text{eq}}$, where $T _{\text{eq}}$ is the temperature of the visible
sector at matter-radiation equality. This condition is required for
successful LSS formation (e.g. \cite{volkaspetraki}). The redshift of
matter-radiation equality is $z _{\text{eq}} = 3200 \pm 130$ \cite{pdg},
which leads to a lower limit on the matter-radiation equality
temperature of about $T _{\text{eq}} = 0.72$ eV.

We can now scan the parameter space of this model and set constraints on
its parameters. In principle the model presents five parameters : $m
_{F_1}$, $m
_{F_2}$, $\alpha ^{'}$, $\epsilon$ and $Z ^{'}$. A numerical analysis of
the solution, $T _{\text{dr}} \gtrsim T _{\text{eq}}$, 
shows a weak dependence on $m _{F_2}$. This can be understood by noting
that an iterative solution of Eq.(\ref{sahafinal}) displays a log-like
dependence on the value of the constant $A$, which is the only place
where $m _{F_2}$ comes into play. The dependence on $Z ^{'}$ is also
relatively minor, since as previously noted it only affects the binding
energy in a modest way.

To summarize, the physics of dark recombination, to a rough
approximation, depends on just 3 parameters: $m _{F_1}$, $\alpha ^{'}$
and $\epsilon$ (being relatively insensitive to $Z'$ and $m _{F_2}$). We
now derive constraints on these 3 parameters.

\vskip 0.8cm
\begin{figure}[htpb]
    \centering
        \includegraphics[scale=0.5, angle=270]{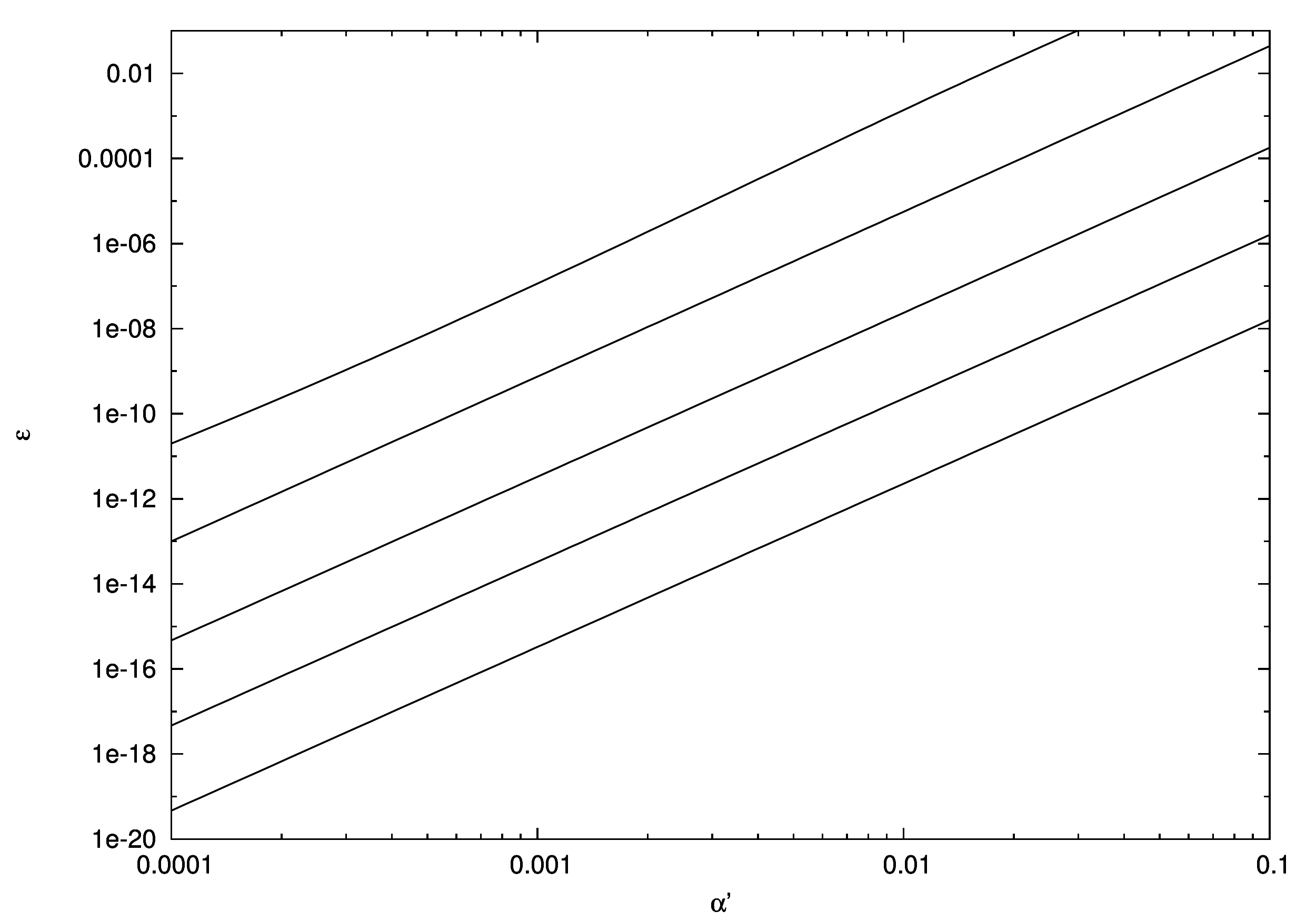}
    \caption{Exclusion limits from the constraint on the temperature of
dark recombination (discussed in text). The limits are for fixed values
of $m _{F_1}$ for (going from upper to lower line) $m _{F_1} = 100, 10,
1, 0.1, 0.01$ MeV (excluded region is above the line).}
    \label{fig:Exclusion dark recombination mf1}
\vskip 0.8cm
\end{figure}

As already discussed, we derive exclusion limits by requiring $T
_{\text{dr}} \gtrsim T _{\text{eq}}$, and using
Eqs.(\ref{sahafinal},\ref{tdark},\ref{darkord},\ref{scaling}) [we take
$I = 10 \ {\rm eV}$ in Eq.(\ref{scaling})]. In Figure \ref{fig:Exclusion
dark recombination mf1} we give the results for a fixed $m _{F_1}$ and
varying $\alpha ^{'}$. The dependence on $\alpha ^{'}$, $m _{F_1}$ shown
in Figure \ref{fig:Exclusion dark recombination mf1} can be easily
understood by analytical considerations. Recall, to constrain the model
we look for the value of parameters for which $T _{\text{dr}} \gtrsim T
_{\text{eq}} \simeq 0.72$ eV. From Eqs.(\ref{tdark},\ref{scaling}) we
have that $T ^{'} \propto I ^{'} \propto {\alpha '} ^2 m _{F_1}$, while
Eq.({\ref{masterformula}) implies $T _{\text{dr}} = T _{\text{dr}} ^{'}T
_{\text{dr}}/T _{\text{dr}} ^{'} \propto {\alpha '} ^2m _{F_1}{\cal M}
^{\frac{1}{4}}/\sqrt{\epsilon}$. It follows therefore that the upper
limit on $\epsilon$ scales as ${\alpha '} ^4 m _{F_1} ^2\sqrt{\cal M}$.
In fact, the numerical results shown in Figure {\ref{fig:Exclusion dark
recombination mf1} give the upper bound on $\epsilon$, coming from dark
recombination:
\begin{eqnarray}
\epsilon \lesssim 10 ^{-8} \left ( \frac{\alpha '}{\alpha} \right ) ^4
\left ( \frac{m _{F_1}}{\rm MeV} \right ) ^2 \left ( \frac{{\cal M}}{m
_e} \right ) ^{\frac{1}{2}} \ .
\label{boundse}
\end{eqnarray}
The above upper bound also includes a factor of $\sim$ 2 uncertainty on
$\epsilon$ arising from the uncertainty on $I$ [$I = (10 \pm 5) \ {\rm
eV}$].

\section{Galactic structure}

In this section we explore small-scale phenomenology of this dissipative
dark matter model, focussing on the structure of spiral and irregular
galaxies
at the present epoch.
In these galaxies the dark matter halo is (currently) assumed to be in
the form of a dissipative plasma
composed of $F_1$ and $F_2$ particles. Such a plasma can be
approximately spherical and extended even in the presence
of substantial energy loss due to dissipative processes (such as dark
bremsstrahlung) provided there exists
a substantial heat source.
Spiral and irregular galaxies exhibit ongoing star formation making it
possible for ordinary core-collapse
supernovae to be this halo heat source (with the halo having evolved as
a consequence of the assumed dynamics so that the heating and cooling
rates balance). 
This mechanism requires kinetic mixing with $\epsilon \sim 10^{-9}$ to
convert a significant fraction of the supernovae core collapse
energy into the production of light $F_1,\bar F_1$ particles and
ultimately into dark photons.
Here, we provide a fairly simplistic analytic treatment of the problem
adapting and expanding aspects of previous work in the MDM 
context \cite{spheroidal,depth4}.
This will, of course, only represent a zeroth-order
approximation which could be improved in a more sophisticated treatment.
Nevertheless, this simple analytic approach provides useful 
insight and should be adequate for the purposes of extracting
the parameter space region of interest.

We will also briefly consider
elliptical and dwarf spheroidal galaxies.
These galaxies must have a different dark matter structure
to spirals and irregulars (at least at the present epoch)
as these galaxy types have little current star formation activity. We
will briefly comment on how these 
galaxy types might
fit into this picture.
The detailed structure of larger systems such as galaxy clusters is of
course very important but will
be left for future work.

\subsection{Dynamical halo model and halo scaling relations}

The physical picture of spiral galaxies is that of a flat disk of
baryonic matter surrounded by a dark matter halo. In our model, the dark
matter halo is formed by a plasma of $F _1$ and $F _2$ particles, where
energy is lost to dissipative interactions, such as thermal dark
bremsstrahlung. To account for the observed halo structure, a heat
source that can replace this energy lost has to exist. In the MDM
context, it has been argued that ordinary supernovae can supply this
energy \cite{spheroidal,depth4,review}. The mechanism involves kinetic
mixing induced processes ($e\overline{e} \rightarrow e'\overline{e}'$,
$\gamma \rightarrow e'\overline{e}'$,...) in the supernovae core, which
can convert $\sim 1/2$ of the core collapse energy into $\gamma ^{'}$,
$e'$, $\overline{e}'$ for $\epsilon \sim 10 ^{-9}$ \cite{raf,updated}
(see also \cite{silagadze}). Ultimately this energy is reprocessed into
mirror photons in the region around the supernovae. Essentially the same
mechanism can take place in our generic two-component dissipative dark
matter model provided that $m _{F_1} \lesssim {\rm few} \times T
_{\text{SN}} \approx 100 \ {\rm MeV}$.


The physical properties of the dark matter halo are then governed by the
Euler equations of fluid dynamics, which take the form:
\begin{gather}
\frac{\partial \rho}{\partial t} + \nabla \cdot (\rho \mathbf{v}) = 0
\nonumber \ , \\
\frac{\partial \mathbf{v}}{\partial t} + (\mathbf{v} \cdot
\nabla)\mathbf{v} = -\left ( \nabla \phi + \frac{\nabla P}{\rho} \right)
\nonumber \ , \\
\frac{\partial}{\partial t} \left [\rho \left ( \frac{\mathbf{v}^2}{2} +
{\cal E} \right ) \right ] + \nabla \cdot \left [\rho \left (
\frac{\mathbf{v}^2}{2} + \frac{P}{\rho} + {\cal E} \right ) \mathbf{v}
\right ] - \rho \mathbf{v} \cdot \nabla \phi = \frac{d\Gamma
_{\text{heat}}}{dV} - \frac{d\Gamma _{\text{cool}}}{dV} \ .
\label{euler}
\end{gather}
Here $P$, $\rho$ and $\mathbf{v}$ denote the pressure, mass density and
velocity of the fluid. ${\cal E}$ is the internal energy per unit mass
of the fluid, so that $\rho \left (\mathbf{v}^2/2 + {\cal E} \right )$
is the energy per unit volume. Finally, $\Gamma _{\text{heat}}$ and
$\Gamma _{\text{cool}}$ are the heating and cooling rates. Significant
simplifications occur if the system evolves to a static configuration.
In this limit, all time derivatives in Eqs.(\ref{euler}) vanish, and if
one also assumes spherical symmetry,\footnote{
For the most part we assume spherical symmetry. This is a simplifying
approximation 
which we expect will lead to reasonable zeroth order results.
Of course, the halo cannot be exactly spherically symmetric; deviations
from spherical symmetry
might be important and future work could attempt to incorporate these.  
Two main sources of asymmetry are the supernova heat source,
distributed within the 
galactic disk, and possible bulk rotation of the halo.
The latter effect depends on the size of the halos angular momentum,
which is unknown and may be difficult to estimate
reliably from theoretical considerations.
} 
then Eqs.(\ref{euler}) reduce to just two equations:
\begin{eqnarray}
\frac{d\Gamma _{\text{heat}}(r)}{dV} = \frac{d\Gamma
_{\text{cool}}(r)}{dV} \ ,
\label{balance}
\end{eqnarray}
and
\begin{eqnarray}
\frac{dP(r)}{dr} = -\rho (r)g(r) \ .
\label{hydrostatic}
\end{eqnarray}
Here $g(r)=\nabla \phi$ is local gravitational acceleration.

The quantities $g(r)$, $P(r)$ can be related to the density $\rho (r)$
via:
\begin{eqnarray}
g(r) &=& \frac{v _{\text{rot}} ^2}{r} = \frac{G}{r ^2} \int _0 ^r dr' \
4\pi {r'} ^2 \rho _T (r') \ , \nonumber \\
P(r) &=& \frac{\rho(r)T(r)}{\overline{m}} \ ,
\label{grhop}
\end{eqnarray}
where 
we have assumed local thermal equilibrium in order to relate $P$ to $T$ and
$\overline{m}$ 
is the mean mass of the particles forming the dark plasma.
[$\overline{m} = (n _{F_1}m _{F_1} + n _{F_2}m _{F_2})/(n _{F_1} + n _{F_2})$ 
for a fully ionized plasma.]
Here $\rho _T (r)$ is the total mass density which, in addition to the
dark plasma component, $\rho (r)$, includes baryonic components (stars
and gas) and possibly compact dark ``stars".

A few comments on Eqs.(\ref{balance},\ref{hydrostatic}) are in order.
Eq.(\ref{balance}) represents energy balance at every point in the halo,
while Eq.(\ref{hydrostatic}) is the hydrostatic equilibrium condition.
Both conditions are required for a static configuration. Whether or not
the system is able to evolve to such a static configuration is not
certain, but seems possible. Assuming that the system, at an early time
prior to the onset of ordinary star formation ($t \lesssim {\rm few} \
{\rm Gyr}$), was in a more compact configuration, then the subsequent
star formation activity would expand and heat the halo (that is, $\Gamma
_{\text{heat}} - \Gamma _{\text{cool}} > 0$ initially), which in turn
would modify $\Gamma _{\text{heat}}-\Gamma _{\text{cool}}$ via various
feedback processes. The idea is that these feedback processes can reduce
$\Gamma _{\text{heat}} - \Gamma _{\text{cool}}$ as the halo expands
until $\Gamma _{\text{heat}} - \Gamma _{\text{cool}} = 0$ is reached.
For example, as the halo expands, the ordinary supernovae rate reduces
in response to the weakening gravity, as expressed by the
Schmidt-Kennicutt empirical law, which relates star formation rate to
the gas density in spiral galaxies \cite{schmidt}:
\begin{eqnarray}
\dot{\Sigma} _{\star} \propto n _{\text{gas}} ^N \ , \ N \sim 1\text{-}2
\ .
\end{eqnarray}
This mechanism and others can potentially lead to a net reduction in
$\Gamma _{\text{heat}} - \Gamma _{\text{cool}}$ as the halo expands,
until eventually the static limit is reached where $\Gamma_{\text{heat}}
= \Gamma_{\text{cool}}$.

In order to gain insight, we initially solve Eq.(\ref{hydrostatic})
assuming an isothermal halo, i.e. $dT/dr = 0$, and approximating $\rho
_T (r) = \rho (r)$. Both of these approximations can be roughly valid in
the outer regions of the galaxy. Combining
Eqs.(\ref{hydrostatic},\ref{grhop}) and taking into account the
isothermal approximation, the hydrostatic equilibrium equation can be
expressed as:
\begin{eqnarray}
\frac{d\rho}{dr} = -\frac{\overline{m}\rho (r) G}{T r ^2}\int _0 ^r dr'
\ 4\pi {r'} ^2 \rho (r') \ .
\label{hydrostatic1}
\end{eqnarray}
Eq.(\ref{hydrostatic1}) can be solved by a polynomial of the form $\rho
= \lambda/r ^p$. Substitution into Eq.(\ref{hydrostatic1}) yields $p =
2$ and $\lambda = T/2\pi G\overline{m}$, that is:
\begin{eqnarray}
\rho (r) &=& \frac{T}{2\pi G\overline{m}r ^2} \ .
\label{ntrho}
\end{eqnarray}
Combining Eqs.(\ref{grhop},\ref{ntrho}) gives us the rotational velocity
profile, which we can relate to the temperature of the halo:
\begin{eqnarray}
v _{\text{rot}} ^2 = \frac{G}{r}\int _0 ^r dr' \ 4\pi {r'}
^2\frac{T}{2\pi G\overline{m}{r'} ^2} = \frac{2T}{\overline{m}} \implies
T = \frac{1}{2}\overline{m}v _{\text{rot}} ^2 \equiv
\frac{1}{2}\overline{m}v _{\infty} ^2 \ .
\label{vrot}
\end{eqnarray}
The rotational velocity is found to be independent from the distance to
the center of the galaxy, consistent with the observed asymptotically
flat rotational curves of spiral galaxies, with asymptotic velocity $v
_{\infty}$.

\subsubsection{Toy model}

Is the assumption of an isothermal halo justified? Let us consider a toy
model, where we consider all supernovae as acting as a point source at
the galactic centre ($r = 0$) producing a total dark photon luminosity
$L _{\text{SN}}$. Clearly this model is unphysical, and will have to be
refined later. To apply Eq.(\ref{balance}) to the system, we have to
match the energies absorbed and dissipated within a volume element $dV$.

Supernovae are presumed to be a source of dark photons, resulting from
kinetic mixing induced processes (e.g. $\gamma \rightarrow
F_1\overline{F}_1$, $e\overline{e} \rightarrow F_1\overline{F}_1$)
occurring in the supernovae cores. The resulting interactions in the
region around the supernovae convert this energy into dark photons of
uncertain spectrum. These dark photons can eventually escape and
ultimately transport and inject the energy into the halo. Two possible
mechanisms can be 
envisaged: dark photoionization and dark Thomson scattering. We show in
Appendix B that dark Thomson scattering is an unimportant heating
mechanism for the
parameter space we are focussing on ($m _{F_1} \gtrsim 0.1 \ \rm MeV$).

Assuming, then, that the heating of the halo takes place via a dark
photoionization process with cross-section $\sigma _{_{DP}}$, the energy
per unit time being absorbed in a given volume element, $dV$, is given
by:\footnote{In principle one has to integrate over the frequency
spectrum of dark photons, as in \cite{depth4}, but this detail is not
essential for the current discussion.}
\begin{eqnarray}
d\Gamma _{\text{heat}} = \frac{L _{\text{SN}}e ^{-\tau}}{4\pi r
^2}\sigma _{_{DP}}n _{F_2} dV \ ,
\label{ein}
\end{eqnarray}
where $\tau$ is the optical depth. We have assumed that the two K-shell
atomic $F_1$ states are occupied, which means that the plasma cannot be
completely ionized. We shall here assume that the remaining $(|Z'|-2)$
$F_1$ states are free, and will comment more on these consistency
conditions in Section 5.2.2. Evidently, the validity of our model then
requires $|Z'| \geq 3$.

Energy is lost via dark bremsstrahlung of $F _1$ off $F _2$. The energy
dissipated per unit time within a volume element $dV$ is given by:
\begin{eqnarray}
d\Gamma _{\text{cool}} = \Lambda (T) n _{F_1}n _{F_2} dV \ ,
\label{eout}
\end{eqnarray}
where $\Lambda(T)$ is the cooling function for dark bremsstrahlung
(defined more precisely in Section 5.2) and $n _{F_1}$ (henceforth)
denotes the free $F_1$ particles number density. There are other sources
of dissipation,
such as line emission and recombination, which
could be included by modifying $\Lambda$ (see e.g. \cite{radiative}).
\footnote{One could also consider inverse Compton
scattering, $F_1 \gamma _{_D} \rightarrow F_1 \gamma _{_D}$, where
$\gamma _{_D}$ is a dark microwave background photon. For the range of
parameter space and physical conditions we are examining, we find that
inverse Compton scattering can be neglected except possibly at an early
epoch, $z \gtrsim 3$.} 
Although they might be important, for the purposes of this discussion
they will be neglected.\footnote{A more comprehensive discussion of
cooling would have to take into account the cooling efficiency. In
general not all bremsstrahlung dark photons will have mean free path
sufficiently long as to escape the halo. Whether or not they can escape
(and hence cool) the halo depends on their location of production and
their wavelength. These effects could be incorporated by means of a
cooling efficiency function which depends on these variables. However,
such a discussion is beyond the scope of our paper and will be left for
future work.} Matching of heating and cooling corresponds to equating
the right-hand sides of Eqs.(\ref{ein},\ref{eout}), which yields:
\begin{eqnarray}
n _{F_1} = \frac{L _{\text{SN}}e ^{-\tau}}{\Lambda (T)4\pi r ^2}\sigma
_{_{DP}} \ .
\label{match}
\end{eqnarray}
If, in addition, we make the assumption that the halo is optically thin
($\tau \ll 1$), we recover $n _{F_1} \propto 1/r ^2$. This also means
that $\rho \propto 1/r ^2$.

The end result is that the assumption of an isothermal halo provides a
solution to both energy balance [Eq.(\ref{balance})] and the hydrostatic
equilibrium condition [Eq.(\ref{hydrostatic})]. This suggests that an
isothermal halo can be a reasonable approximation at large distances
from the galactic centre, where the supernova heat source can be
modelled as a point source and where, in addition, $\rho _T (r) \simeq
\rho (r)$.

\subsubsection{A refined model: solution to the core-cusp problem}

The toy model described above is unphysical at $r = 0$. To refine it, we
smear the supernova energy source over a finite volume, on a distance
scale $r_D$. Since we are dealing with ordinary supernovae, it is
reasonable to assume they are distributed similarly to the mass of the
galactic disk. One therefore expects the $\rho \propto 1/r ^2$ solution
to hold only for $r \gg r _D$. The mass distribution of the galactic
disk can be approximated by a profile known as Freeman disk, with
surface density \cite{freeman}:
\begin{eqnarray}
\Sigma (\widetilde{r}) = \frac{m _D}{2\pi r_D ^2}e
^{-\frac{\widetilde{r}}{r_D}} \ ,
\end{eqnarray}
with $r_D$ being the disk scale length and $m _D$ its total mass. 

We can now follow the same steps as in \cite{review}. Using cylindrical
coordinates ($\widetilde{r},\widetilde{\theta},\widetilde{z}$) and
setting the disk at $\widetilde{z} = 0$, the flux at a point $P = (r
_1,0,z _1)$ within an optically thin halo is given by:
\begin{eqnarray}
f(r,\cos \phi ) = \frac{L _{\text{SN}}}{4\pi m _D} \int
d\widetilde{\theta} \int d\widetilde{r} \ \widetilde{r} \frac{\Sigma
(\widetilde{r})}{{\widetilde{r}} ^2 - 2\widetilde{r}r _1 \cos
\widetilde{\theta} + r _1 ^2 + z _1 ^2}
\end{eqnarray}
where $r = \sqrt{r _1 ^2 + z _1 ^2}$ and $\cos \phi \equiv r _1/r$. It
is not hard to show that:
\begin{eqnarray}
f(r,\cos \phi) \propto \begin{cases}
                  \log r, & r \lesssim r_D \ , \\
                  \frac{1}{r ^2}, & r \gg r_D \ .
                  \end{cases}
\label{frc}
\end{eqnarray}
The energy lost per unit time due to thermal dark bremsstrahlung is once
again given by Eq.(\ref{eout}), while the energy absorbed per unit time
within a volume element $dV$ now takes the form:
\begin{eqnarray}
d\Gamma _{\text{heat}} = f(r,\cos \phi)\sigma _{_{DP}}n _{F _2}dV \ .
\label{ein1}
\end{eqnarray}
Again equating d$\Gamma _{\text{heat}}$=d$\Gamma _{\text{cool}}$, using
Eqs.(\ref{eout},\ref{ein1}), implies $n _{F_1} = f(r,\cos \phi)\sigma
_{_{DP}}/\Lambda (T)$. That is, $\rho \propto f(r,\cos \phi )$.

\begin{figure}[htpb]
    \vskip 1cm
    \centering
        \includegraphics[scale=0.5, angle=270]{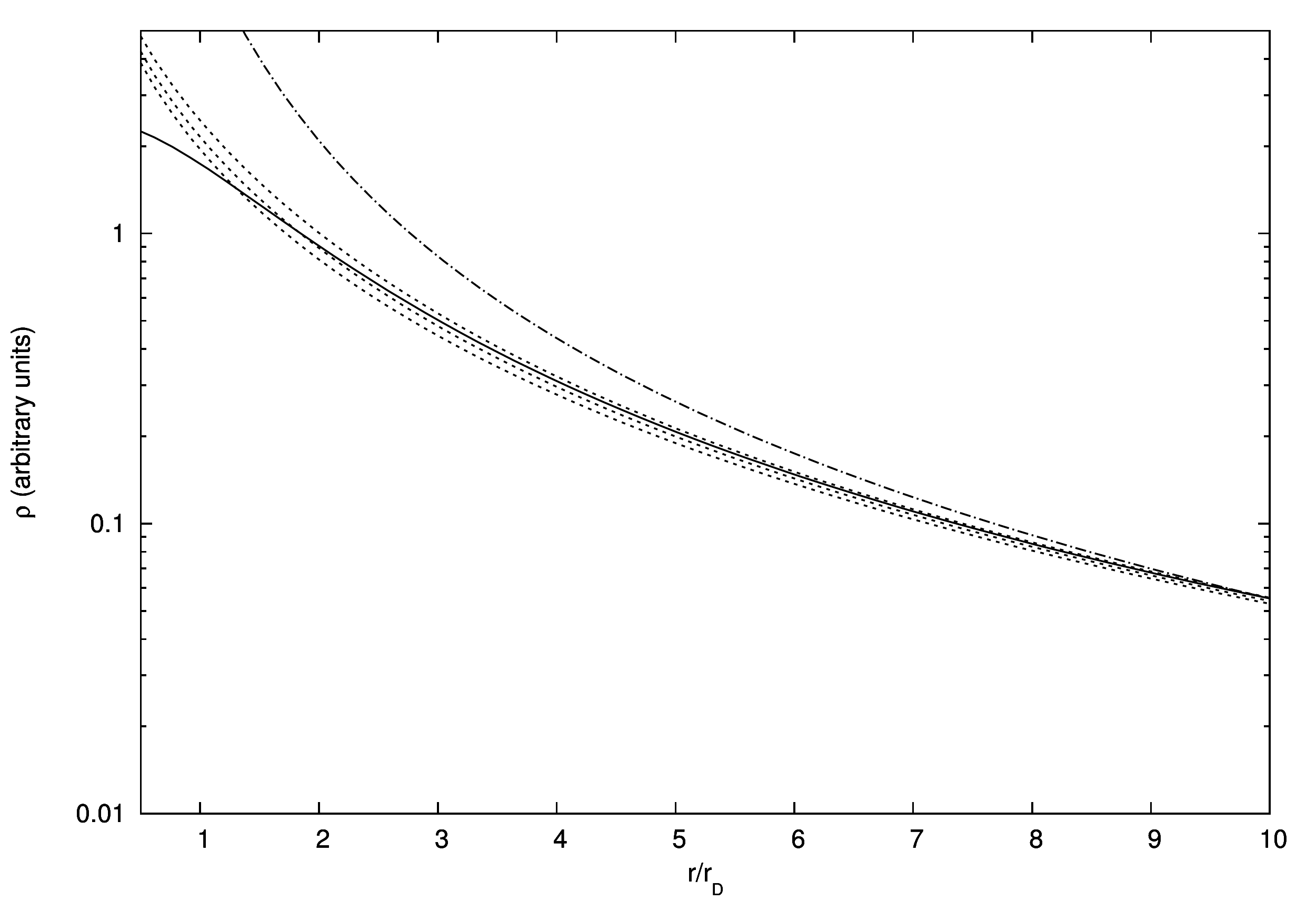}
    \caption{Comparison between the radial dependence of $\rho \propto
f(r,\cos \phi)$, 
the quasi-isothermal profile given by Eq.(\ref{core}), and a cuspy
profile $\rho \propto 1/r ^2$ (in arbitrary units). 
The dotted lines correspond to $f(r,\cos \phi)$ for (going from upper to
lower line) $\phi=\pi/4,\pi/3,\pi/2$. The solid line corresponds to a
cored density profile (with $r _0/r _D = 1.4$), while the dot-dashed
line corresponds to the cuspy profile.}
    \label{fig:Comparison core f}
    \vskip 0.8cm
\end{figure}

The above considerations 
regarding the behavior of $f(r,\cos \phi)$ [Eq.(\ref{frc})] then suggest
that $\rho (r)$ can be approximated by a quasi-isothermal dark matter
profile:
\begin{eqnarray}
\rho (r) \simeq \frac{\rho _0r _0 ^2}{r ^2 + r _0 ^2} \ ,
\label{core}
\end{eqnarray}
where $r _0 \sim r_D$, since the latter is the only length scale present
in the problem. 
In Figure \ref{fig:Comparison core f} we compare the radial dependence
of the 
solution $\rho \propto f(r,\cos \phi)$ with the quasi-isothermal profile
given by Eq.(\ref{core}), finding 
good agreement up to $r \simeq r _D$. 
[Differences at low radii, $r \stackrel{<}{\sim} r_D$, are not so important as baryons typically
dominate the matter density in this region.]

Note that the dark matter density profile obtained in Eq.(\ref{core}) is
cored rather than 
cuspy (as would be if $\rho \propto 1/r ^2$), with the cored profile
arising from having smeared the supernova energy source 
over a finite volume. This suggests a simple explanation for the
inferred existence of dark matter cores in disk galaxies. The 
inability to explain the cored dark matter profile is one of the
shortcomings of collisionless CDM, and is referred 
to as the core-cusp problem (for a review see e.g. \cite{deblok}). In
addition, the scaling relation $r _0 \sim r_D$ 
is actually implied by measurements of high resolution rotation curves
\cite{donato}:
\begin{eqnarray}
\log r _0 = (1.05 \pm 0.11) \log r_D + (0.33 \pm 0.04) \ .
\end{eqnarray}
Eq.(\ref{core}) and the scaling relation $r _0 \sim r _D$ have been
derived by considering energy balance within a given galaxy. 
There is another piece of information we have yet to utilize. That is,
demanding that the total energy input must 
match the total energy output for every disk galaxy.

\subsubsection{Tully-Fisher relation}

If the system evolves to a static configuration, where the heating and
cooling rates  balance, then the properties of galactic halos will be
constrained. Moreover, since heating is proportional to the supernovae
rate and cooling is related to the properties of dark matter, energy
balance will imply a connection between the baryonic and dark matter
components in spiral galaxies. The heating rate of the halo in a given
spiral galaxy can be expressed as:
\begin{eqnarray}
\Gamma _{\text{heat}} = f _{\text{SN}}\langle E _{\text{SN}} \rangle R
_{\text{SN}} \ ,
\label{tullyfisherein}
\end{eqnarray}
where $E _{\text{SN}}$ is the total energy output from each supernova,
and $R _{\text{SN}}$ is the rate at which supernovae occur. The fraction
of energy which is absorbed by the halo, $f _{\text{SN}}$, is given by:
\begin{eqnarray}
f _{\text{SN}} = R _{\gamma _{_D}}\langle \left (1 - e ^{-\tau} \right
)\rangle \ ,
\label{fsn}
\end{eqnarray}
where the fraction of the total energy output in dark particles is $R
_{\gamma _{_D}} \equiv E _D/E _{\text{SN}}$, $E _D$ being the amount of
energy released from the supernova which is ultimately converted into
the creation of dark photons. As a measure of the average optical depth,
we consider dark photons propagating from the galactic centre to the
edge of the galaxy (approximated as $r \rightarrow \infty$):
\begin{eqnarray}
\tau = \int _0 ^{\infty} dr \ \sigma _{_{DP}}n _{F_2} = \int _0
^{\infty} dr \ \sigma _{_{DP}} \rho \kappa = \frac{\pi \sigma _{_{DP}}
\kappa \rho _0r _0}{2} \ ,
\label{opticaldepth}
\end{eqnarray}
where we have made use of the density profile given in Eq.(\ref{core})
and related the density to the $F _2$ number density via:
\begin{eqnarray}
\rho = n _{F_2}(m _{F_2} + |Z'|m _{F_1}) \equiv \frac{n _{F_2}}{\kappa}
\ .
\label{kappa}
\end{eqnarray}
Combining Eqs.(\ref{tullyfisherein}-\ref{kappa}) it follows that, in the
optically thin limit, the heating rate for the halo of a given spiral
galaxy is:
\begin{eqnarray}
\Gamma _{\text{heat}} = \frac{\pi R _{\gamma _{_D}}\sigma _{_{DP}}\kappa
\langle E _{\text{SN}} \rangle}{2} \ \rho _0r _0R _{\text{SN}} \ .
\label{tullyfisherein1}
\end{eqnarray}

The differential cooling rate of the halo is given by Eq.(\ref{eout}).
To obtain the total cooling rate Eq.(\ref{eout}) has to be integrated in
the volume element. In doing so, note that the differential cooling rate
depends on the parameters defining the dark matter density profile,
$\rho _0$ and $r _0$, through $n _{F _1} \approx \rho \kappa (|Z'|-2)$
and $n _{F_2} = \rho \kappa$.\footnote{The relation $n _{F_1} \approx
\rho \kappa (|Z'|-2)$ assumes the plasma is not fully ionized, but has
the K-shell states occupied, so that dark photoionization can occur.
More generally, $n _{F_1} = f\rho \kappa (|Z'|-2)$, where $f \leq 1$
accounts for partial ionization of the remaining atomic states.}
Integrating Eq.(\ref{eout}) yields:
\begin{eqnarray}
\Gamma _{\text{cool}} = \Lambda (T)\kappa ^2 (|Z'|-2)\rho _0 ^2r _0
^4\int _0 ^{\infty} dr' \ \frac{4\pi {r'} ^2}{({r'} ^2 + r _0 ^2) ^2} =
\pi ^2\kappa ^2(|Z'|-2)\Lambda (T)\rho _0 ^2r _0 ^3 \ .
\label{tullyfishereout}
\end{eqnarray}
Under the assumption that the main source of dissipation is thermal dark
bremsstrahlung, $\Lambda (T) \propto \sqrt{T}$ (see e.g.
\cite{radiative}). The temperature $T$ is related to the rotational
velocity of the galaxy far from the center, $v _{\infty}$, via
Eq.(\ref{vrot}), so that $\Lambda (T) \propto \sqrt{T} \propto v
_{\infty}$. The rotational velocity profile (having neglected baryonic
contributions), $v _{\text{rot}}(r)$, can be related to $\rho _0$ and $r
_0$ via Eq.(\ref{grhop}):
\begin{eqnarray}
v _{\text{rot}} ^2 = \frac{G}{r} \int _0 ^r dr' \ 4\pi {r'} ^2
\frac{\rho _0r _0 ^2}{{r'} ^2 + r _0 ^2} = 4\pi G\rho _0 r _0 ^2 \left
[1 - \frac{r _0}{r}\tan ^{-1} \left ( \frac{r}{r _0} \right ) \right ] \
.
\label{vrot1}
\end{eqnarray}
For $r \gg r _D$, we then have:
\begin{eqnarray}
v _{\infty} = \left (4\pi G\rho _0r _0 ^2 \right) ^{\frac{1}{2}} \ .
\label{vrot2}
\end{eqnarray}

Imposing the energy balance condition [Eq.(\ref{balance})], and hence
equating $\Gamma _{\text{heat}} = \Gamma _{\text{cool}}$, with $\Gamma
_{\text{heat}}$ and $\Gamma _{\text{cool}}$ given by
Eqs.(\ref{tullyfisherein1},\ref{tullyfishereout}), we find:
\begin{eqnarray}
\Lambda (T)\rho _0 r _0 ^2 = \frac{R _{\gamma _{_D}}\sigma
_{_{DP}}\langle E _{\text{SN}} \rangle}{2\pi \kappa (|Z'|-2)} \ R
_{\text{SN}} \ .
\label{tullyf}
\end{eqnarray}
This represents a scaling relation connecting dark matter properties
($\rho _0$ and $ r_0$) with baryonic properties, such as $R
_{\text{SN}}$ (and is independent of the previously obtained $r _0 \sim
r _D$ relation). We show below that it is roughly equivalent to the
empirical Tully-Fisher relation. Combining
Eqs.(\ref{vrot2},\ref{tullyf}) and recalling that $\Lambda (T) \propto v
_{\infty} \propto \left ( \rho _0 r _0 ^2 \right ) ^{\frac{1}{2}}$
results in a scaling relation connecting the supernovae rate and the
asymptotic rotational velocity in a given spiral galaxy:
\begin{eqnarray}
R _{\text{SN}} \propto v _{\infty} ^3 \ .
\label{tullyfisher1}
\end{eqnarray}
Supernovae observational studies have found the relation $R _{\text{SN}}
\propto \left ( L _B \right ) ^{0.73}$ \cite{wli}, where $L _B$ is the
galaxy $B$-band luminosity. Combining this relation with that in
Eq.(\ref{tullyfisher1}) yields:
\begin{eqnarray}
L _B \propto v _{\infty} ^4 \ .
\label{tullyfisher}
\end{eqnarray}
Eq.(\ref{tullyfisher}) is one of the forms of the Tully-Fisher relation
(see e.g. \cite{webster}), an empirical 
relation that is observed to hold for spiral galaxies \cite{tully} and
used extensively as a rung on the 
cosmic distance ladder (see for instance \cite{maoz}). The general form
of the Tully-Fisher relation is $L \propto \left ( v _{\text{rot}}
\right ) ^{\alpha}$, where the power $\alpha$ depends on the luminosity
band under consideration. For instance, for the $K$-band (near-infrared)
$\alpha = 4.35 \pm 0.14$ is determined \cite{webster}, while for the
optical $B$-band $\alpha = 3.91 \pm 0.13$ is found \cite{webster}. The
Tully-Fisher relation is currently unexplained, although it suggests a
deep connection between the baryonic and dark matter components of
spiral galaxies. Our model seems to supply such a connection via the
nontrivial dissipative dynamics: the Tully-Fisher relation is the energy
balance condition, Eq.(\ref{balance}), where $\Gamma _{\text{heat}}$
arises from supernovae heating and $\Gamma _{\text{cool}}$ from
dissipative dynamics.\footnote{It is worth mentioning that a third
relation, not independent from the other two ($r _0 \propto r _D$ and $L
_B \propto v _{\infty} ^4 \propto \rho _0 ^2r _0 ^4$), can be obtained.
Observational studies have shown that $m _D \propto \left ( L _B \right
) ^{1.3}$ \cite{shankar} and $r_D \propto \left ( m _D \right ) ^{0.38}$
\cite{saluccird}. Combining these relations yields $\rho _0 r _0 \approx
{\rm constant}$ (which is observed to hold in spiral galaxies
\cite{kormendy}).} This scenario is expected to hold within irregular
galaxies as well, since these galaxies have ongoing star formation like
spirals.

\subsubsection{Elliptical galaxies: the Faber-Jackson relation}

The dynamical halo model, with heating powered by kinetic mixing induced
processes in the core of ordinary supernovae 
balancing the energy loss due to dissipative processes in the halo,
seems to be
viable for galaxies with ongoing star formation: that is, spiral and
irregular galaxies. This picture cannot be directly applied to
elliptical galaxies 
or dwarf spheroidal galaxies
as these galaxies are devoid of baryonic gas and exhibit suppressed star
formation. 
Focussing first on ellipticals (we briefly discuss dwarf spheroidals in
the following subsection)
it is possible that these galaxies could have evolved from spirals. In
particular, spirals may have a final evolutionary stage where 
they have exhausted their baryonic gas to the point where the ordinary
supernova rate is insufficient to support the dark halo from collapse.

Consider the limiting case where $t _{\text{cool}} \ll t _{\text{ff}}$,
with $t _{\text{cool}}$ and $t _{\text{ff}}$ being the cooling and
free-fall timescales respectively. In this limit, the dark halo can cool
and potentially fragment into dark stars. Imagine a point in time where
the heating suddenly stops and the halo cools but does not have time yet
to collapse (consistently with $t _{\text{cool}} \ll t _{\text{ff}}$).
The total energy at this time can be approximated as just the
gravitational potential energy, and is given by:
\begin{eqnarray}
U _i = -\int _0 ^R dr \ 4\pi r^2 \frac{GM_r}{r}\rho (r) \ ,
\label{ui}
\end{eqnarray}
where the mass enclosed within a radius $r$ is:
\begin{eqnarray}
M_r = \int _0 ^r dr' \ 4\pi {r'} ^2\rho (r') \simeq 4\pi \rho _0r _0 ^2
r \ .
\end{eqnarray}
In evaluating $M_r$ above, we have used the density profile given by
Eq.(\ref{core}).  In the limit where $t _{\text{cool}} \ll t
_{\text{ff}}$, this should be a good approximation, as the dark matter
density profile has no ``time" to change. Evaluating the integral in
Eq.(\ref{ui}) then gives:
\begin{eqnarray}
U _i = -4\pi \rho _0r _0 ^2 GM _t \ ,
\end{eqnarray}
where the total mass is $M _t = \int _0 ^R dr \ 4\pi r ^2 \rho \simeq
4\pi \rho _0r _0 ^2R$. 

As the system contracts, and assuming dark stars form, these stars would
attain kinetic energy as they fall into the gravitational potential
well. The virial theorem can then be used to relate their eventual
kinetic energy, in terms of the eventual potential energy: $U _f = -2T
_f$. Thus, equating this final energy with the initial energy gives:
\begin{eqnarray}
U _i = U _f + T _f = -T _f \ ,
\end{eqnarray}
By using $T _f = 3M _t\sigma _v ^2/2$, where $\sigma_v$ is the average
velocity dispersion of the dark stars, we find that:
\begin{eqnarray}
\sigma _v ^2 = \frac{8\pi G\rho _0r _0 ^2}{3} \ .
\end{eqnarray}
If, in addition, we make the assumption that the ordinary stars
``thermalize" with the dark stars, it follows that their velocity
dispersion will also be approximately $\sigma _v ^2$. Given that the
elliptical galaxy in the picture evolved from a spiral galaxy, the
$\rho_0, r_0$ parameters obey the scaling relations derived earlier.
Using the scaling relation $\rho _0r _0 \approx$ constant and $r _0
\propto r _D \propto \sqrt{L _B}$, which follows from $m _D \propto
\left ( L _B \right ) ^{1.3}$ \cite{shankar} and $r_D \propto \left ( m
_D \right ) ^{0.38}$ \cite{saluccird}, we obtain a relation between the
$B$-band luminosity of a given elliptical galaxy and its velocity
dispersion:
\begin{eqnarray}
L _B \propto \sigma _v ^4 \ .
\end{eqnarray}
Such a scaling relation, known as the Faber-Jackson relation
\cite{faberjackson}, is observed to roughly hold for elliptical
galaxies.

This picture of elliptical galaxies might help explain some of their
distinctive properties. In particular, if the dark stars produce dark
supernovae then kinetic mixing induced processes in the core of these
dark supernovae can generate a large flux of ionizing ordinary photons,
which can heat ordinary matter, thereby potentially explaining why
elliptical galaxies are observed to be devoid of baryonic gas.

\subsubsection{Dwarf spheroidal galaxies}

Dwarf spheroidal galaxies, like ellipticals, are also devoid of baryonic
gas and show little star formation activity (at the present epoch).
It is possible that they reach this point in their evolution in a manner
broadly analogous to the picture just described above for ellipticals
(although their formation may have been very different).
That is, at an earlier stage in their evolution 
these galaxies had a dark matter plasma halo which had dynamically
evolved into a steady state configuration
featuring hydrostatic equilibrium and with heating and cooling rates
balanced.
Then at some point, perhaps due to insufficient star formation to keep
up with the heating 
requirements, the halo collapsed and fragmented into
dark stars. If this dark star formation rate is rapid enough the dark
matter structural properties of the galaxy can be preserved.
In this manner it might be possible to explain why dwarf spheroidal
galaxies, irregular/spirals, and ellipticals all have 
broadly similar dark matter structual properties as indicated from
observations 
(e.g. the inferred dark matter surface density, $\rho_0 r_0$, is roughly
constant independent of galaxy type \cite{donato2}).

Although the middle and latest stages in the evolution of dwarf
spheroidal and elliptical galaxies might be similar (as discussed
above),
their formation may have been very different. 
Studies of the dwarf spheroidal population around Andromeda (M31) galaxy
show that a large fraction of these satellites orbit in a thin plane
\cite{plan1}. 
(A similar planar structure of satellites, although not quite so
impressive, has also been observed around the Milky Way \cite{plan2}).
These
observations can potentially be explained if the dwarf spheroidal
galaxies formed during a major merger event, so that they are in fact
tidal dwarf galaxies \cite{Kroupa}. Even if a significant fraction of
dwarf spheroidal galaxies formed in this way, 
they can still be dark matter dominated and have evolved via the
dissipative dynamics 
so that their current structural properties are consistent 
with observations (e.g. with scaling relations such as the roughly
constant dark matter surface density, $\rho_0 r_0$).
At the earliest stages of galaxy formation, prior to ordinary star
formation, the dark matter which seeded
the galaxy may have collapsed into a disk due to the dissipative
processes.
Subsequently the ordinary baryons also formed a disk. Gravitational
interactions between the two disks
can cause them to merge on a fairly short time scale cf.\cite{dddm}.
A major galaxy merger event around this time could have produced tidal
dwarf galaxies with large dark matter fraction
(as the dark matter particles in the disk have velocities correlated
with the baryonic particles).
The observed alignment of the satellite galaxies around M31
can thereby be potentially explained, 
as was discussed for the mirror dark matter case \cite{footsil}.
Of course, the formation of the ordinary disk and consequent ordinary
star generation and supernovae will lead 
to the production of dark photons (via kinetic mixing induced
processes).
This energy is presumed to eventually heat and expand the disk dark gas
component of the host galaxy (in this case M31)
into its current state: a roughly spherical halo.

\subsection{Consistency conditions and energy balance}

The assumption that the system evolves to a static configuration has
allowed us to establish a connection between the baryonic and dark
matter components in disk galaxies, in the form of scaling relations
which are consistent with observations. We now wish to understand how
this energy balance argument can constrain the 5-dimensional parameter
space of our dark matter model. This requires a more quantitative
understanding of the exact heating and cooling mechanisms.

As previously discussed, thermal dark bremsstrahlung of $F _1$ off $F
_2$ is assumed to be the dominant dissipation avenue. The energy lost
per unit time per unit volume due to this process is given in e.g.
\cite{radiative}:
\begin{eqnarray}
\frac{d\Gamma _{\text{cool}}}{dV} = \frac{16 {\alpha '} ^3 (2\pi T)
^{\frac{1}{2}}}{(3 m _{F_1} ) ^{\frac{3}{2}}} {Z '} ^2 n _{F_1} n _{F_2}
\overline{g} _B \ ,
\label{bremsstrahlung}
\end{eqnarray}
where $\overline{g} _B \simeq 1.2$ is the frequency average of the
velocity-averaged Gaunt factor for thermal bremsstrahlung. 

The temperature, $T$, in Eq.(\ref{bremsstrahlung}), is related to the
mean mass of the dark plasma. In the limit where $m _{F_2} \gg m
_{F_1}$, and assuming the two K-shell atomic states are occupied, neutrality of the plasma implies
that the number density of free $F_1$ states is: $n _{F_1} =
(|Z'|-2)n _{F_2}$. In this circumstance the mean mass can be approximated by:
\begin{eqnarray}
\overline{m} = \frac{n _{F_1}m _{F_1} + n _{F_2}m _{F_2}}{n _{F_1} + n
_{F_2}} \approx \frac{m _{F_2}}{|Z'|-1}
\ .
\label{meanmass}
\end{eqnarray}
Using
Eqs.(\ref{vrot},\ref{core},\ref{kappa},\ref{vrot2},\ref{bremsstrahlung},\ref{meanmass}),
the total cooling rate can be expressed as:
\begin{eqnarray}
\Gamma _{\text{cool}} = 32\pi ^3\overline{g} _B{\alpha '} ^3{Z'}
^2(|Z'|-2)\kappa ^2\sqrt{\frac{Gm _{F_2}}{27(|Z'|-1)m _{F_1} ^3}}\left (
\rho _0r _0 \right) ^{\frac{5}{2}}r _0 ^{\frac{3}{2}} \ .
\label{integratedout}
\end{eqnarray}

In the MDM framework it has been argued that photoionization of K-shell
mirror electrons in a mirror metal component can replace the energy lost
due to dissipation. This process can take place because the mirror
metals in question retain their K-shell mirror electrons \cite{review}.
If we assume that in our model $D ^0$ (the dark bound state), albeit
being close to fully ionized, retains its K-shell $F _1$ particles, then
a similar mechanism, which we call \textit{dark photoionization}, can
efficiently heat the halo. The cross-section for dark photoionization,
$\sigma _{_{DP}}$, can be easily obtained from that of ordinary
photoionization, found in e.g. \cite{radiative}:
\begin{eqnarray}
\sigma _{_{DP}} = \frac{g'16\sqrt{2}\pi}{3 {m _{F_1}} ^2} {\alpha '} ^6
{|Z'|} ^5 \left ( \frac{m _{F_1}}{E _{\gamma _{_D}}} \right )
^{\frac{7}{2}} \ ,
\label{photoionization}
\end{eqnarray}
where $g' = 1,2$ counts the number of K-shell $F_1$ particles present.

For the picture we have just presented to be valid, a series of
consistency conditions will have to hold. We will now proceed to discuss
what these conditions are and how they constrain the available parameter
space for our model.

\subsubsection{Cooling timescale}

The dynamical halo picture, governed by a balance between heating and
cooling rates, could only hold provided the cooling timescale is much
less than the Hubble time. This requirement constrains the available
parameter space and, as one can see from Eq.(\ref{integratedout}), will
set an upper bound on the mass of $F_2$ [recall $\kappa ^{-1} = (m
_{F_2} + |Z'|m _{F_1}$)]. If $n _T$ is the total number density of dark
particles, the cooling timescale is given by:
\begin{eqnarray}
t _{\text{cool}} \approx \frac{\frac{3}{2}n _TT}{\Lambda (T)n _{F_1}n
_{F_2}} \approx \frac{3T}{2\Lambda (T)n _{F_2}} \ ,
\label{cooling}
\end{eqnarray}
where we have approximated $n _T \approx n _{F_1}$. Making use of
Eqs.(\ref{vrot},\ref{kappa},\ref{bremsstrahlung},\ref{meanmass}), we can
write:
\begin{eqnarray}
t _{\text{cool}} \approx \frac{9\sqrt{3}}{64\overline{g}
_B\sqrt{\pi}}\sqrt{\frac{m _{F_2}m _{F_1} ^3}{|Z'|-1}}\frac{v
_{\text{rot}}}{\kappa \rho {\alpha '} ^3{Z'} ^2} \ .
\end{eqnarray}
Observe that the cooling timescale can be defined locally, i.e. $t
_{\text{cool}}(r)$, through the dependence on $\rho (r)$. Less dense
regions cool more slowly, so the most stringent limit occurs where $\rho
(r)$ is lowest. Of course we have little knowledge about halo properties
far from the galactic center. As a rough limit, we shall require  $t
_{\text{cool}}(r) \lesssim$ few billion years, for $r \lesssim 3.2 r _D
\sim 2 r _0$ (3.2 $r _D$ is the optical 
radius where most of the baryons reside, defined in e.g.
\cite{optical}). Note that the most stringent limits occur for the
largest disk galaxies, where $\rho (r=2r _0) \approx \rho _0/5$ and $v
_{\text{rot}} \approx 300 \ \rm km/s$. Here we have taken the typical
values (for large disk galaxies) $\rho _0r _0 \simeq 100 \ \rm M
_{\astrosun}/pc ^2$ and $r _0 \simeq 20 \ \rm kpc$ and hence $\rho _0/5
\simeq 10 ^{-3} \ {\rm M _{\astrosun}/pc ^3}$. In this case, the
requirement $t _{\text{cool}} (r=2r _0) \lesssim$ few billion years
gives the upper limit on the mass of $F_2$:
\begin{eqnarray}
m _{F_2} \lesssim 200 \left ( \frac{\rm MeV}{m _{F_1}} \right )\left (
\frac{\alpha '}{10 ^{-2}} \right ) ^2\left ( \frac{|Z'|}{10} \right )
^{\frac{5}{3}} \ \rm GeV \ .
\label{mf2cool}
\end{eqnarray}

\subsubsection{Ionization state of the halo}

The scenario described earlier assumed that the halo is ionized but the
dark bound state, $D ^0$, retains its K-shell $F_1$ particles. The
former requirement allows for efficient cooling via dark bremsstrahlung,
while the latter is a necessary condition for dark photoionization to
take place. Here we require such a picture to hold for all disk
galaxies, regardless of size. Were this not the case, one would expect
significant observational differences in moving along the spectrum of
disk galaxies, depending on whether or not their dark plasma is ionized
or $D ^0$ retains its K-shell $F_1$ particles. Hence we require the
temperature of the halo, given in Eq.(\ref{vrot}), to be high enough to
ensure that $D ^0 $ is ionized (at least one free $F_1$ particle per
bound state), while being low enough as to allow the K-shell $F_1$
particles be retained. 

By comparing the appropriate ionization and capture cross-sections, in
Appendix A we estimate that, given the ionization energy ${\cal I}$, the
transition from an ionized to a neutral halo occurs at a temperature $T
= {\cal I}/\xi$, where $\xi \approx 7-28$. Of course, in the process of
obtaining a conservative lower bound on the mass of the $F_2$ particle,
we are interested in the maximum value $\xi$ can assume, that is, $\xi
_{\max} \approx 28$. Similarly, to obtain a conservative upper bound on
$m _{F_2}$, we are interested in the minimum value $\xi$ can assume in
relation to the process of K-shell photoionization. In Appendix A we
estimate that $\xi _{\min} \approx \min [1/({\alpha '} ^3{Z'} ^4),1]$,
and hence, denoting by ${\cal J}$ the relevant ionization energy, we
obtain the rough conditions:
\begin{eqnarray}
T \gtrsim \frac{{\cal I}}{\xi _{\max}} \implies \frac{m _{F_2}}{\rm GeV}
& \gtrsim & \left (\frac{|Z'|}{10} \right )\left (\frac{\alpha '}{10
^{-2}} \right ) ^2\left ( \frac{m _{F_1}}{\rm MeV} \right )\left (
\frac{50 \ {\rm km/s}}{v _{\text{rot}}} \right ) ^2 \ , \nonumber \\
T \lesssim \frac{{\cal J}}{\xi _{\min}} \implies \frac{m _{F_2}}{\rm
GeV} & \lesssim & 100 \left ( \frac{|Z'|}{10} \right ) ^3\left
(\frac{\alpha '}{10 ^{-2}} \right ) ^2\left ( \frac{m _{F_1}}{\rm MeV}
\right )\left ( \frac{300 \ {\rm km/s}}{v _{\text{rot}}} \right )
^2g(\alpha ',Z') \ ,
\nonumber \\
~  
\label{mf2ij}
\end{eqnarray}
where $g(\alpha ',Z') \equiv \max({\alpha '} ^3{Z'} ^4,1)$. Clearly the
most stringent lower bound on $m _{F_2}$ arises from the smallest
spiral/irregular galaxies, with $v _{\text{rot}} \approx 50 \ \rm km/s$,
while the most stringent upper bound comes from the biggest disk
galaxies, for which $v _{\text{rot}} \approx 300 \ \rm km/s$, and thus:
\begin{eqnarray}
\left (\frac{|Z'|}{10} \right )\left (\frac{\alpha '}{10 ^{-2}} \right )
^2\left ( \frac{m _{F_1}}{\rm MeV} \right ) \lesssim \frac{m
_{F_2}}{{\rm GeV}} \lesssim 100 \left ( \frac{|Z'|}{10} \right ) ^3\left
(\frac{\alpha '}{10 ^{-2}} \right ) ^2\left ( \frac{m _{F_1}}{\rm MeV}
\right )g(\alpha ',Z')
\label{mf2bounds}
\end{eqnarray}
In addition, we have to require that the upper bound on $m _{F_2}$
[Eq.(\ref{mf2cool})] be greater than the respective lower bound
[Eqs.(\ref{mf2ij})]. Doing so yields:
\begin{eqnarray}
|Z'| \gtrsim 4 \left ( \frac{m _{F_1}}{10 \ \rm MeV} \right ) ^3 \ .
\label{boundz}
\end{eqnarray}

It is conceivable that the ionization physics sets the physical scale of
spiral/irregular galaxies (i.e. sets either or both $v _{\text{rot}}
^{\max}$, $v _{\text{rot}} ^{\min}$), which means that either or both
the limits in Eq.(\ref{mf2bounds}) are equalities. Equating the two
bounds in Eq.(\ref{mf2bounds}) we obtain that this limiting situation
occurs for $|Z'| \sim 1$.

\subsubsection{Energy balance}

We now turn to the energy balance condition, $\Gamma _{\text{heat}} =
\Gamma _{\text{cool}}$ [Eq.(\ref{balance})] As previously discussed, we
have assumed that the galactic system evolves such that this condition
is currently satisfied for disk galaxies. Given the observed properties
of disk galaxies, we can use this condition to constrain the fundamental
parameters of our model.

The cooling rate, assuming the main dissipation process being dark
bremsstrahlung, is readily found [Eq.(\ref{integratedout})]. For the
heating rate the situation is more complicated. Details about $\Gamma
_{\text{heat}}$ require a detailed understanding of the frequency
spectrum of the dark photons which, it is alleged, heat the halo.
Nevertheless, we can set an upper limit on the value of this heating
rate:
\begin{eqnarray}
\Gamma _{\text{heat}} \lesssim R _{\gamma _{_D}}R _{\text{SN}}\langle E
_{\text{SN}} \rangle \min(\tau _{\max},1) \ ,
\label{gammaheat}
\end{eqnarray}
where $\tau _{\max}$ is the maximum value the optical depth
[Eq.(\ref{opticaldepth})] can take after allowing for all possible forms
of the $\gamma _{_D}$ spectrum.
Eqs.(\ref{opticaldepth},\ref{photoionization}) suggest that the optical
depth is maximized when $E _{\gamma _{_D}} = I'$, where $I' \approx {Z'}
^2{\alpha '} ^2m _{F_1}/2$ is the ionization energy of the relevant
K-shell $F_1$ particle, hence:
\begin{eqnarray}
\tau _{\max} = \frac{256\pi ^2}{3}\frac{\rho _0r _0}{m _{F_1} ^2m
_{F_2}\alpha '{Z'} ^2} \ .
\label{taumax}
\end{eqnarray}
Assuming the nominal value $\rho _0r _0 \approx 100 \ M
_{\astrosun}/{\rm pc ^2} \simeq 4.6 \times 10 ^{-6} \ {\rm GeV ^3}$ and
taking the upper bound on $m _{F_2}$ given in Eq.(\ref{mf2bounds}), we
get:
\begin{eqnarray}
\tau _{\max} \gtrsim 40 \left ( \frac{\rm MeV}{m _{F_1}} \right )
^3\left ( \frac{10 ^{-2}}{\alpha '} \right ) ^3\left ( \frac{10}{|Z'|}
\right ) ^5\frac{1}{g(\alpha ',Z')} \ .
\label{tau}
\end{eqnarray}
Eq.(\ref{tau}) suggests $\tau _{\max}\gtrsim 1$ holds for a significant
fraction of parameter space.

Let us now assume parameters where $\tau _{\max} \gtrsim 1$ and evaluate
an upper limit for $\Gamma _{\text{heat}}$ [note that even with
parameters where $\tau _{\max} \lesssim 1$, the derived limit will be
still valid, given that $\min (\tau _{\max},1) \leq 1$]. For $\epsilon
\lesssim 10 ^{-9}$, $R _{\gamma _{_D}} \propto \epsilon ^2$, while for
$\epsilon \gtrsim 10 ^{-9}$, $R _{\gamma _{_D}}$ actually saturates at
$\sim 1/2$. By inserting numbers into Eq.(\ref{gammaheat}), we get:
\begin{eqnarray}
\Gamma _{\text{heat}} \lesssim \left (\frac{\epsilon}{10 ^{-9}} \right )
^2\left (\frac{\langle E _{\text{SN}} \rangle}{3 \times 10 ^{53} \ \rm
erg} \right )\left (\frac{R _{\text{SN}}}{0.03 \ \rm yr ^{-1}} \right )
10 ^{44} \ \rm \frac{erg}{s} \ ,
\label{numberheat}
\end{eqnarray}
which holds for $\epsilon \lesssim 10 ^{-9}$. Similarly, inserting
numbers into Eq.(\ref{integratedout}), we obtain the cooling rate for a
given galaxy:
\begin{eqnarray}
\Gamma _{\text{cool}} \simeq \left (\frac{\alpha '}{10 ^{-2}} \right )
^3\left (\frac{\rm MeV}{m _{F_1}} \right ) ^{\frac{3}{2}}\left (
\frac{|Z'|}{10} \right ) ^{\frac{5}{2}}\left (\frac{10 \ \rm GeV}{m
_{F_2}} \right ) ^{\frac{3}{2}}\left (\frac{\rho _0r _0}{100 \ \frac{M
_{\astrosun}}{\rm pc ^2}} \right ) ^{\frac{5}{2}}\left (\frac{r _0}{5 \
\rm kpc} \right ) ^{\frac{3}{2}} 10 ^{44} \ \rm \frac{erg}{s} \ .
\nonumber \\
~ 
\label{numbercool}
\end{eqnarray}
Comparison of Eqs.(\ref{numberheat},\ref{numbercool}) requires the
following approximate relation holds:
\begin{eqnarray}
{\cal C}\left (\frac{10 ^{-9}}{\epsilon} \right ) ^2\left (\frac{\alpha
'}{10 ^{-2}} \right ) ^3\left (\frac{\rm MeV}{m _{F_1}} \right )
^{\frac{3}{2}}\left (\frac{|Z'|}{10} \right ) ^{\frac{5}{2}}\left
(\frac{10 \ \rm GeV}{m _{F_2}} \right ) ^{\frac{3}{2}} \lesssim 1 \ ,
\label{rsn}
\end{eqnarray}
where:
\begin{eqnarray}
{\cal C} \equiv \left (\frac{\rho _0r _0}{100 \ \frac{M
_{\astrosun}}{\rm pc ^2}} \right ) ^{\frac{5}{2}}\left (\frac{r _0}{5 \
\rm kpc} \right ) ^{\frac{3}{2}}\left (\frac{3 \times 10 ^{53} \ \rm
erg}{\langle E _{\text{SN}} \rangle} \right )\left ( \frac{0.03 \ \rm yr
^{-1}}{R _{\text{SN}}} \right ) \ .
\label{calc}
\end{eqnarray}
We expect ${\cal C} \approx 1$ to hold for all spirals on account of
scaling relations. In addition, Eqs.(\ref{mf2cool},\ref{rsn}) provide us
with a
rough lower bound on $\epsilon$:
\begin{eqnarray}
\epsilon \gtrsim 10 ^{-10} \ .
\label{boundepsilon}
\end{eqnarray}
Note that this lower bound is consistent with the upper bounds on
$\epsilon$ derived previously from early Universe cosmology.

\vskip 0.2cm

\section{Summary of the bounds on the model}

\vskip 0.1cm

Having studied the early Universe cosmology and galactic structure
implications of the model, we can now make use of our analyses to
constrain the 5-dimensional parameter space in question. We start by
looking at the kinetic mixing parameter, $\epsilon$. The validity of our
picture of galaxy structure requires core-collapse supernovae to produce
a considerable energy output in light dark particles (specifically,
$F_1\overline{F}_1$ pairs initially) via kinetic mixing induced
processes. We have found that $\epsilon \gtrsim 10 ^{-10}$ is required
for the energy output to successfully heat the halo
[Eq.(\ref{boundepsilon})]. An upper bound on $\epsilon$ was derived in
Section 3 from $\delta N _{\text{eff}}$[CMB] and $\delta N
_{\text{eff}}$[BBN] constraints (Figure \ref{fig:Comparison bbn cmb}),
which indicate $\epsilon \lesssim 5 \times 10 ^{-8}$.

As discussed in Section 2, $m _{F_1}$ is required to be bounded above by
about 100 MeV, otherwise $F _1\overline{F} _1$ pair production becomes
exponentially (Boltzmann) suppressed in the core of core-collapse
supernovae, where the maximum temperature which can be reached is of
about 30 MeV. A lower limit of around $m _{F_1} \gtrsim 0.01 \ \rm MeV$
arises from White Dwarf cooling  and Red Giants helium flash
considerations \cite{updated}.

A constraint on the dark recombination temperature (so that dark
acoustic oscillations do not modify the early growth of LSS) also
provided a useful constraint on parameters. This constraint,
Eq.(\ref{boundse}), together with the above limits on $m _{F_1}$,
$\epsilon$, suggest a lower bound: $\alpha ' \gtrsim 10 ^{-4}$. Further,
our analysis implicitly assumed that perturbation theory could reliably
be used to calculate cross-sections, ionization energies, and so forth,
which is only valid if $\alpha '$ is sufficiently small: $\alpha '
\lesssim 10 ^{-1}$.

Constraints on $m _{F_2}$ were derived from galactic structure
considerations in Section 5. There it was shown that a successful
picture of spiral and irregular galaxies could be achieved within this
two-component hidden sector model provided $m _{F_2}$ satisfies the
constraints given by Eqs.(\ref{mf2cool},\ref{mf2bounds},\ref{rsn}).


Below, we summarize the bounds obtained in this work:

\begin{eqnarray}
\begin{cases}
\epsilon \lesssim \min \left [3.5 \times 10 ^{-9} \left ( \frac{{\cal
M}}{m _e} \right ) ^{\frac{1}{2}} \ , \ 10 ^{-8} \left ( \frac{\alpha
'}{\alpha} \right ) ^4 \left ( \frac{m _{F_1}}{\rm MeV} \right ) ^2
\left ( \frac{{\cal M}}{m _e} \right ) ^{\frac{1}{2}} \right ] \ , \\
\epsilon \gtrsim 10 ^{-10} \ , \\
0.01 \ {\rm MeV} \lesssim m _{F_1} \lesssim 100 \ {\rm MeV} \ , \\
m _{F_2} \gtrsim \left (\frac{|Z'|}{10} \right )\left (\frac{\alpha
'}{10 ^{-2}} \right ) ^2\left ( \frac{m _{F_1}}{\rm MeV} \right ) \ {\rm
GeV} \ , \\
m _{F_2} \lesssim \min \left [200 \left ( \frac{\rm MeV}{m _{F_1}}
\right )\left ( \frac{\alpha '}{10 ^{-2}} \right ) ^2\left (
\frac{|Z'|}{10} \right ) ^{\frac{5}{3}} \ , \ 100 \left (
\frac{|Z'|}{10} \right ) ^3\left (\frac{\alpha '}{10 ^{-2}} \right )
^2\left ( \frac{m _{F_1}}{\rm MeV} \right )g(\alpha ',Z') \right ] \
{\rm GeV} , \\
10 ^{-4} \lesssim \alpha ' \lesssim 10 ^{-1} , \\
|Z'| \gtrsim \max \left [ 3, 4 \left ( \frac{m _{F_1}}{10 \ \rm MeV}
\right ) ^3 \right ] \ ,
\end{cases}
\label{boundsinterrelated}
\end{eqnarray}
where ${\cal M} \equiv \max ( m _e, m _{F_1} )$ and $g (\alpha ',Z')
\equiv \max ({\alpha '} ^3{Z'} ^4,1)$ [$m _e = 0.511 \ {\rm MeV}$ is the
electron mass].

There is a finite, but certainly restricted, region of parameter space
consistent with all of the above constraints. For example, if we fix $m
_{F_1} = 1 \ \rm  MeV$, $\alpha ' = 10 ^{-2}, |Z'|=10$, the above
constraints are all satisfied for $10 ^{-10} \lesssim \epsilon \lesssim
5 \times 10 ^{-9}$ and $1 \ {\rm GeV} \lesssim m _{F_2} \lesssim 100 \
{\rm GeV}$.


\section{Concluding remarks}

Dark matter can be accommodated without modifying known Standard Model
physics by hypothesizing the existence of a hidden sector. 
That is, an additional sector containing particles and forces which
interact with the known Standard Model particle content predominantly
via gravity. 
We have considered a hidden sector containing two stable particles,
$F_1$ and $F_2$, charged under an unbroken $U(1) ^{'}$ gauge symmetry, 
hence featuring dissipative interactions. The associated massless gauge
field, the dark photon, can 
interact via kinetic mixing with the ordinary photon. Our analysis
indicates that such an interaction, of 
strength $\epsilon \sim 10^{-9}$, is required in order to explain
galactic structure. We calculated the effect of this new 
physics on BBN and its contribution to the relativistic energy density
at Hydrogen recombination. Subsequently 
we examined the process of dark recombination, during which neutral dark
states are formed, which is important for 
LSS formation. 

We then analyzed the phenomenology of our model in the context of
galactic structure. 
Focussing on spiral and irregular galaxies, we modelled their halos (at
the current epoch) as a plasma 
composed of dark matter particles, $F_1$ and $F_2$.
This plasma has a substantial on-going energy loss due to dissipative
processes such as dark bremsstrahlung. Kinetic mixing
induced processes in the core of ordinary supernovae can convert a
substantial fraction of the gravitational core-collapse energy 
into dark sector particles (and eventually into dark photons), that
ultimately provides the halo energy which compensates for the
dissipative energy lost.
We found that such a dynamical picture can reproduce 
several observed features of spiral and irregular galaxies, including
the cored density profile and the Tully-Fisher relation. 
We also discussed how elliptical and dwarf spheroidal galaxies might fit
into this framework which we argued has the potential to
explain many of their peculiar features.

The above considerations constrain the five Lagrangian parameters of our
model, as summarized in Eqs.(\ref{boundsinterrelated}). 
Note, in particular, that the kinetic mixing coupling, $\epsilon$, is
constrained to lie within 
the range $10 ^{-10} \lesssim \epsilon \lesssim 5 \times 10 ^{-8}$. A
correct simultaneous explanation of both early Universe cosmology 
and galactic structure typically requires one fermion, $F_1$, to be in
the MeV range (or just below) and the other to be heavier, 
in the GeV (or possibly TeV) range. 

The allowed mass range of the two fermions means they can be, at least
in principle, detected in direct detection experiments. 
Two types of interactions are of particular interest in this context:
$F_1$-electron scattering and $F_2$-nuclei scattering. 
The self-interacting nature of the $F_1$ and $F_2$ particles enhances
the capture rate of these particles within the Earth, 
giving rise to a unique signature: a diurnal modulation in the
interaction rate. Such an effect is expected to be particularly 
evident for experiments located in the Southern hemisphere, giving rise
to suppressions in the interaction rate which could be as large as 100\%
\cite{mine}.

Although an explanation of the DAMA annual modulation signal
\cite{dama1} in terms of nuclear recoils appears disfavored given the
null results of the other experiments, recent work (in the context of
MDM) has shown that it might be possible to explain it in terms of dark
matter scattering off electrons if the mass of the dark matter particle
is in the MeV range \cite{electronscattering}. Within the framework of
our two-component model, a similar explanation seems possible, that is,
the observed annual modulation signal in the DAMA experiment might be
due to $F_1$-electron scattering.

Hidden sector dark matter models can be quite appealing from a
theoretical point of view, and, as we have shown, can provide a
satisfactory explanation for dark matter phenomena on both large and
small scales. In our study we have constrained the parameter space of a
particularly simple two component hidden sector model, and have
indicated potential ways of testing such a model in the context of
direct detection experiments.


\section*{Appendix A}

We estimate the quantity $\xi$ in Section 5.2.2. Recall, $\xi$ is
defined in terms of the transition temperature between two states, at
the relevant ionization energy $I'$, $T = I'/\xi$. Consider, for
instance, the process relevant for $D ^0$ ionization, with cross-section
$\sigma _I$:
\begin{eqnarray}
F_1 + D ^0 \rightarrow D ^+ + F_1 + F_1 \ ,
\label{ionziationd}
\end{eqnarray}
which is opposed by the corresponding capture process, with
cross-section $\sigma _C$:
\begin{eqnarray}
F_1 + D ^+ \rightarrow D ^0 + \gamma _{_D} \ .
\label{captured}
\end{eqnarray}
The number density of $D ^+$ is governed by the following rate equation:
\begin{eqnarray}
\frac{dn _{D ^+}}{dt} = n _{F_1}n _{D ^0}\langle \sigma _Iv
_{F_1}\rangle - n _{F_1}n _{D ^+}\langle \sigma _Cv _{F_1}\rangle \ .
\label{ratedp}
\end{eqnarray}
It follows that in a steady-state situation $n _{D ^+}/n _{D ^0}=\langle
\sigma _Iv _{F_1} \rangle/\langle \sigma _Cv _{F_1} \rangle$, and hence
we compare the relevant thermally averaged ionization and capture
cross-sections:
\begin{eqnarray}
\langle \sigma _Iv _{F_1} \rangle &=& \sqrt{\frac{1}{m _{F_1}\pi}}\left
(\frac{2}{T}\right ) ^{\frac{3}{2}}\int _{I'} ^{\infty} dE _{F_1}E
_{F_1}e ^{-\frac{E _{F_1}}{T}}\sigma _I \ , \nonumber \\
\langle \sigma _Cv _{F_1} \rangle &=& \sqrt{\frac{1}{m _{F_1}\pi}}\left
(\frac{2}{T}\right ) ^{\frac{3}{2}}\int _0 ^{\infty} dE _{F_1}E _{F_1}e
^{-\frac{E _{F_1}}{T}}\sigma _C \ .
\label{sigivsigcv}
\end{eqnarray}
The ionization and capture cross-sections are given in
\cite{lotz,kramers} and are roughly:\footnote{The following expressions
assumes the $F_1$ particles are non-relativistic, that is, $T \lesssim m
_{F_1}$. If we demand that the non-relativistic approximation is valid
for all spirals ($v _{\text{rot}} \lesssim 300 \ \rm km/s$), then we
require $m _{F_2}/m _{F_1} \lesssim 10 ^6(|Z'|-1)$.}
\begin{eqnarray}
\sigma _I & \sim & \frac{{\alpha '} ^2}{E_{F_1}I'} \ , \nonumber \\
\sigma _C & \sim & \frac{{\alpha '} ^5{Z'} ^4}{E _{F_1}(E _{F_1} + I')}
\ .
\end{eqnarray}
The relevant transition will occur when the quantity $\langle \sigma _Iv
_{F_1} \rangle / \langle \sigma _Cv _{F_1} \rangle$ is of order 1, that
is:
\begin{eqnarray}
\frac{\langle \sigma _Iv _{F_1} \rangle}{\langle \sigma _Cv _{F_1}
\rangle} \sim \frac{I'+T}{I'}\frac{e ^{-\frac{I'}{T}}}{{\alpha '} ^3{Z'}
^4} \sim 1 \ ,
\end{eqnarray}
and hence when:
\begin{eqnarray}
\left ( 1+\frac{1}{\xi} \right ) e ^{-\xi} \sim {\alpha '} ^3{Z'} ^4 \ ,
\label{xiexi}
\end{eqnarray}
where $\xi \equiv I/T$.

For the process of $D ^0$ ionization, we can safely take $|Z'| \approx
1$. Solving Eq.(\ref{xiexi}) shows that a value of $\xi \sim 7-28$ is
the solution within the allowed range of parameter space ($10 ^{-4}
\lesssim \alpha ' \lesssim 10 ^{-1}$). In Section 5.2.2,
Eqs.(\ref{mf2ij}), we obtain the most conservative lower bound on the
mass of $F_2$ when $\xi = \xi _{\max} \approx 28$.

When analyzing the process of K-shell dark photoionization,
Eqs.(\ref{mf2ij}), we obtain the most conservative upper bound on the
mass of $F_2$ when $\xi$ assumes its lowest possible value. In this case
we find that, to a reasonable approximation, $\xi _{\min} \approx \min
[1/({\alpha '} ^3{Z'} ^4),1]$.

\section*{Appendix B}

In the paper we assumed that the dark photons arising from kinetic
mixing induced processes in the core of ordinary supernovae heat the
halo via a dark photoionization process. In principle, one could
consider dark Thomson scattering ($\gamma _{_D}F_1 \rightarrow \gamma
_{_D}F_1$, where $F_1$ denotes a free $F_1$ particle) as an equally
viable heating mechanism. However, we will show below that this is not
expected to be the case for the parameter space of interest.

The optical depth for dark Thomson scattering, considering a dark photon
propagating from the center of the galaxy to infinity, is given by:
\begin{eqnarray}
\tau = \int _0 ^{\infty} dr \ \sigma _{_{DT}}n _{F_1} = \int _0
^{\infty} dr \ \sigma _{_{DT}}\rho \kappa (|Z'|-2) = \frac{4\pi
^2{\alpha '} ^2\kappa(|Z'|-2)\rho _0r _0}{3m _{F_1} ^2} \ ,
\label{opticaldepthdark}
\end{eqnarray}
where we have related the free $F_1$ number density to the density
profile via the relation $n _{F_1} \approx \rho \kappa (|Z'|-2)$ and
made use of the expression for the dark Thomson scattering cross-section
$\sigma _T = 8\pi{\alpha '} ^2/(3m _{F_1} ^2)$.

Assuming the spectrum of dark photons that heat the halo has energy
spectrum that peaks well below the electron mass, kinematic
considerations dictate that dark Thomson scattering can only efficiently
impart energy to the scattered $F_1$ particles provided that $\tau \gg
1$ (i.e. the dark photon becomes trapped within the galaxy), and hence
if:
\begin{eqnarray}
m _{F_2} \ll \frac{4\pi ^2\rho _0r _0}{3}\frac{{\alpha '} ^2|Z'|}{m
_{F_1} ^2} \ .
\label{mf2thomson}
\end{eqnarray}
Here we have used $\kappa \approx 1/m _{F_2}$ [from Eq.(\ref{kappa})].
Recall, the basic requirement that the halo be ionized gave a lower
bound on $m _{F_2}$ [Eqs.(\ref{mf2ij})]. Requiring that the above upper
bound on $m _{F_2}$ [Eq.(\ref{mf2thomson})] be greater than the lower
bound found in Eqs.(\ref{mf2ij}), we find:
\begin{eqnarray}
m _{F_1} ^3 \ll \frac{4\pi ^2\rho _0r _0v _{\text{rot}} ^2\xi
_{\max}}{3} \ .
\label{mf1thomson}
\end{eqnarray}
Eq.(\ref{mf1thomson}) reduces to:
\begin{eqnarray}
\frac{m _{F_1}}{{\rm MeV}} \ll \left ( \frac{\rho _0r _0}{100 \ \frac{M
_{\astrosun}}{{\rm pc ^2}}} \right ) ^{\frac{1}{3}}\left ( \frac{v
_{\text{rot}}}{300 \ \rm km/s} \right )^{\frac{2}{3}} \ .
\end{eqnarray}
This is the condition for dark Thomson scattering to be a viable heating
mechanism. It follows that dark Thomson scattering is not expected to be
an important
heating mechanism for any spirals ($v _{\text{rot}} \lesssim 300 \ \rm
km/s$) if $m _{F_1} \gtrsim 0.1 \ \rm MeV$, which is the parameter range
we are focussing on.

\vskip 1.7cm

\begin{flushleft}

{\Large \bf Acknowledgments}

\end{flushleft}

\vskip 0.2cm
\noindent
SV would like to thank Rachel Webster, Harry Quiney, Valter Moretti and
Alexander Millar for useful discussions. SV would also like to thank
Jackson Clarke and Brian Le for valuable help on the computational side
of this work. RF would like to thank Alexander Spencer-Smith
for useful correspondence. This work was partly supported by the
Australian Research Council and the Melbourne Graduate School of
Science.


\end{document}